\documentclass[journal]{IEEEtran}
\usepackage{cite}
\usepackage{amsmath}
\usepackage{algorithmic}
\usepackage{array}
\usepackage{graphicx}
\usepackage[dvipsnames]{xcolor}
\usepackage{float}%
\usepackage{multirow}
\usepackage{booktabs}
\usepackage{soul}
\usepackage{microtype}
\usepackage{multirow}

\usepackage{tabularx}
\usepackage[flushleft]{threeparttable}
\usepackage{makecell}

\begin{document}

\title{Retrofitting FSO Systems in Existing RF Infrastructure: A Non-Zero Sum Game Technology}

\author{Abderrahmen Trichili, \textit{Member IEEE}, Amr Ragheb, \textit{Member, IEEE}, Dmitrii Briantcev, \textit{Student Member, IEEE}, Maged A. Esmail, \textit{Member, IEEE}, Majid Altamimi, \textit{Member, IEEE}, Islam Ashry, \textit{Senior Member, IEEE}, Boon S. Ooi, \textit{Senior Member}, IEEE, Saleh Alshebeili, Mohamed-Slim Alouini, \textit{Fellow, IEEE}

\thanks{A. Trichili, D. Briantcev, I. Ashry, B. S. Ooi, and M.-S Alouini are with the Computer, Electrical and Mathematical Sciences $\&$ Engineering in King Abdullah University of Science and Technology. Email: \{abderrahmen.trichili,dmitrii.briantcev,islam.ashry,boon.ooi,slim.alouini\}@
kaust.edu.sa.\newline 
A. Ragheb, M. Altamimi, S. Alshebeili are with King Saud University. Email: \{a.ragheb,mtamimi,dsaleh\}@ksu.edu.sa \newline
M. A. Esmail is with Prince Sultan University. Email: mesmail@psu.edu.sa. }
}


\maketitle
\begin{abstract}
Progress in optical wireless communication (OWC) has unleashed the potential to transmit data in an ultra-fast manner without incurring large investments and bulk infrastructure. OWC includes wireless data transmissions in three optical sub-bands; ultraviolet, visible, and infrared. This paper discusses installing infrared OWC, known as free space optics (FSO), systems on top of installed radio frequency (RF) networks for outdoor applications to benefit from the reliability of RF links and the unlicensed broad optical spectrum, and the large data rates carried by laser beams propagating in free space. We equally review commercially available solutions and the hardware requirements for RF and FSO technology co-existence. The potential of hybrid RF/FSO for space communication is further discussed. Finally, open problems and future research directions are presented.
\end{abstract}


\IEEEpeerreviewmaketitle

\section{Introduction}
\IEEEPARstart{I}{nternet} and data traffic drastically increased over the last few years with the emergence of new online applications such as virtual reality and ultra-high-definition video streaming. Satisfying the demand of the bandwidth-hungry world requires reliable connectivity. Free space optics (FSO) is a license-free technology that connects two communicating terminals using narrow infrared (IR) light beams and has received massive attention over the last few years. FSO can be a solution to the so-called `last mile' and `last meter' connectivity problems and can be used when optical fiber installation is scarce (See Fig.~\ref{fig:FSOUseCases}). FSO has also been proposed for 5G backhauling \cite{FSOBackhauling,ChenCommag20} and expected to play a significant role to build next-generation 6G networks \cite{6GVTMag,6GNatureElectro}. It has been shown that FSO can benefit from spatial mode multiplexing to push the communication capacity by using the spatial structure of the light \cite{TrichiliConst19} and through other various degrees of freedom of light such as the frequency \cite{FSOWDM} and polarization \cite{CvijeticPDM10}. Due to the spectrum scarcity of radio frequency (RF) technology, FSO has been widely proposed as an alternative to existing RF links. Such an idea attracted several negative arguments in particular because of FSO signals' sensitivity to propagation effects. But what if FSO is used as a complementary technology to RF? For instance, RF and FSO links are affected differently with weather conditions \cite{KimSpie11}. FSO links, for example, are sensitive to fog but not to rain, while RF links are severally degraded by rain but resilient to fog. Therefore, RF can be used when the FSO is not available, and FSO can be used when the RF is not available. Furthermore, FSO systems are easily integrable and can be up and running over a short period \cite{FiberWithoutFiber}. More importantly, installing FSO systems does not incur additional civil engineering costs and can be installed on the well-established RF infrastructure. Indeed, RF and FSO technologies' complementary behavior has led to various proposals of hybrid RF/FSO implementations \cite{DouikIEEETCOM16}. In this context, we focus on the various hybrid RF/FSO links. In particular, we present the different RF/FSO link switching mechanisms, such as soft switching and hard switching. The scenario when the RF and FSO systems transmit simultaneously is also reviewed. The practicality of each of these approaches and the hardware challenges of building heterogeneous RF/FSO systems are discussed. Commercially available RF/FSO solutions are equally reviewed. The potentials of the co-existence of RF and FSO technologies for near-Earth and deep space applications are also presented.\\
\subsection{Related Reviews}
Free space optics has been a topic of interest of many published reviews \cite{UysalComst14, GhassemlooyJSAC15, Space_FSO, TrichiliJOSAB20}. From a communication theory perspective, authors of \cite{UysalComst14} reviewed FSO communication and covered the various aspects of FSO links, including channel modeling, modulation techniques, and transceivers' structures. Hybrid RF/FSO systems were briefly discussed, illustrating their potential in combating propagation effects. Ghassemlooy \textit{et al.} highlighted the potential of optical wireless communication systems to complement existing RF networks \cite{GhassemlooyJSAC15}. Authors of \cite{Space_FSO} conducted an extensive survey on optical communication in space and the various challenges and corresponding mitigation strategies. In particular, RF/FSO was presented as a potential technique to mitigate atmospheric turbulence effects. Authors of \cite{OWHybridNetworks} discussed hybrid optical networks; in particular, they overviewed RF/FSO links and their potential to achieve high data rate reliable wireless transmissions. A broad overview of many of the essential topics required to design and develop next-generation FSO systems was provided in \cite{TrichiliJOSAB20}. Retrofitting FSO in existing RF installation was proposed to increase the data rates, and communication reliability of wireless backhauls \cite{TrichiliJOSAB20}. Integrating wireless fidelity (WiFi) and visible light communication (VLC), also known as light fidelity (LiFi), for indoor environments was addressed in \cite{WiFiLiFi}. \newline
\subsection{Contributions}\indent Here, we provide a comprehensive review on integrating IR FSO systems on existing RF infrastructures. We start by giving a brief comparison between RF and FSO to highlight the complementarity of both technologies.  We shed light on the various switching techniques between FSO and RF in a hybrid system. Among these system configurations, we review the contributions in the following approaches:
\begin{itemize}
    \item Soft switching
    \item Hard switching
    \item Machine-learning-based switching
    \item Diversity scheme
\end{itemize}
We also compare them in terms of hardware complexity. The specifications of RF/FSO systems available in the market, field trials, and seminal deployments are covered. Furthermore, the contribution of hybrid RF/FSO for space and near-Earth applications is presented. Issues mainly related to the cost are also discussed. A set of insightful future research directions is finally proposed.
\subsection{Paper Organization}
\indent The remainder of this paper is organized as follows. Section \ref{RFvsFSO} provides a comparison between FSO and RF technologies mainly in terms of licensing, safety, system foot-print, and possible propagation effects. Section \ref{sec:AsymmetricRFFSO} is dedicated to the mixed RF/FSO links. Section~\ref{sec:HybridRFFSO} is devoted to the various configurations of hybrid RF/FSO links. Section \ref{Sec:Hardware} describes commercially available RF/FSO solutions and the main field demonstrations. The potentials of hybrid RF/FSO systems for satellite and deep space applications are discussed in Section~\ref{Sec:SpaceRFFSO}. Open issues and potentials research directions from the authors' perspective are provided on \ref{sec:Direction}. Section~\ref{Conclusion} concludes the paper.
 \begin{figure*}[tb]
        \centering
        \includegraphics[width=\textwidth]{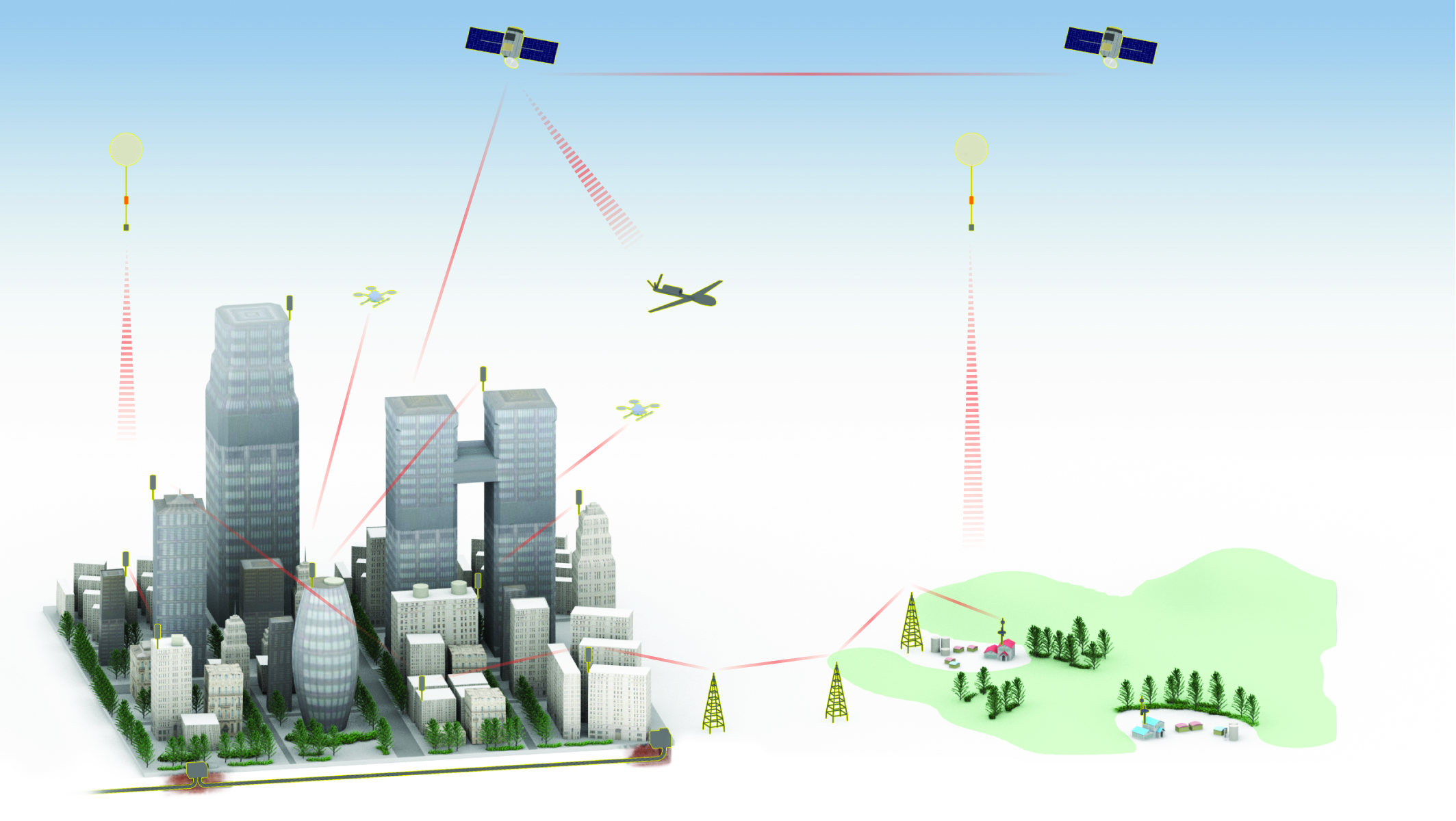}
        \caption{Illustration of potential outdoor FSO use cases ranging from point-to-point terrestrial communication to satellite-to satellite and satellite-to ground connections. \copyright King Abdullah University of Science and Technology.}
        \label{fig:FSOUseCases}
    \end{figure*}

\section{FSO vs RF}
FSO and RF communications, each of which has many advantages and constraints. For example, dust and fog drastically affect FSO, while rain severely affects RF. RF beams tend to diverge higher than FSO ones; however, FSO signals are more sensitive to pointing errors. This section discusses the advantages and disadvantages of FSO and RF systems in terms of bandwidth, footprint, safety, secrecy, and resilience to channel propagation effects. Such discussion will motivate the integration of both technologies to create robust hybrid systems.   
\label{RFvsFSO}
\subsection{Unlicensed Spectrum and Available Bandwidth}
Currently, RF is the technology of choice for wireless communication. However, radio communication requires paying for licenses to use a few bands of the strictly regulated spectrum. FSO, on the other hand, does not suffer from such a problem since the spectrum of light is unlicensed.
In terms of bandwidth, FSO has a vast bandwidth in the range of a few hundred THz, while RF has a much smaller available bandwidth. RF frequencies below 10 GHz are almost exhausted because of their favorable communication properties, while moving to higher frequencies comes with a higher cost, more complexity, and propagation challenges. Current RF technology supports data rates in the range of Mbps/Gbps. However, state-of-the-art FSO intensity modulation/direct detection (IM/DD) technology can support Gbps data rates and reach beyond Tbps when coherent optical systems are used \cite{TrichiliJOSAB20}. We note that in an IM/DD system, the intensity of a laser is modulated using an RF signal. The modulated optical signal is detected using a photodetector at the receiver.   In contrast, a coherent system involves using the intensity and phase of an optical carrier, which requires continuous phase tracking utilizing a device known as a local oscillator. This drive the system complexity and cost higher.\newline
High-speed FSO systems are possible using all-optical networks as light signals coming from an optical fiber can be directed to an FSO channel and vice versa without the need for intermediate electrical to optical and optical to electrical conversion devices that limit the bandwidth. Moreover, this arrangement contributes to reducing the end-to-end system latency.  
 
\subsection{System Footprint and Geometrical Loss}
The desired beamwidth of the received signal after the propagation over a wireless channel primarily defines the receiving terminal size. A narrow beamwidth reduces the amount of power loss and terminal size. The beamwidth, at a position of the receiver $z_{R{X}}$ along the propagation is proportional to the wavelength, $\lambda$, and is inversely proportional to the transmitter aperture diameter $D_T$, that is:
\begin{equation}
\omega (z_{RX}) \sim \frac{\lambda}{D_T}. 
\end{equation}
With a fixed $D_T$, a transition from RF to FSO will significantly reduce the beamwidth $\omega$. For the sake of comparison, let's assume an RF frequency of 50 GHz for a possible 5G system with a corresponding $\lambda_1 \approx 6 \cdot 10^{-3} \text{ [m]}$. Also, consider an infrared  wavelength used for an FSO system with $\lambda_2 = 1550 \text{ [nm]} = 1.55 \cdot 10^{-6} \text{ [m]}$. Even in this, favorable for the RF system comparison scenario, a relative reduction of a spot size coming from RF to FSO will be equal to $\frac{\lambda_1}{\lambda_2} \approx 4 \cdot 10^{3}$. This improvement in the beam waist of more than 3 orders of magnitude reduces the transmitter aperture diameter $D_T$ for a reduced footprint of the FSO system compared to RF. The reduction in the transmitter aperture diameter reduces the signal's power loss at the receiver due to the geometric path loss. The small beam sizes make FSO preferable for backhaul applications, while RF is preferred for access networks. \newline

\subsection{Safety and Secrecy}
FSO has fewer restrictions than the RF technology in terms of power levels and EMF (electromagnetic field) radiation risks because of its point-to-point topology and narrow beam nature \cite{RFHealthRisks}. Similarly, for the same reasons, FSO's physical layer security (PLS) is higher as it is not easy to intercept collimated IR signals using power meters and spectrum analyzers. In addition to being invisible to the naked eye, some infrared FSO signals are also invisible to standard photography, or state-of-the-art video cameras \cite{fsonaSecrecy}. FSO signals cannot penetrate opaque objects along the propagation path, such as walls; therefore, they are secure from hidden adversaries. On the contrary, RF is susceptible to eavesdropping issues. It has also been shown that FSO links can contribute to RF's physical security enhancement when installed parallel to an RF link \cite{SecrecyRF-FSO}.
  The advanced security of wireless IR optical beams comes with a cost that FSO, when compared with RF, is restricted to a point-to-point network configuration, limiting the coverage and reach of such a technology. Note that in some situations, FSO signals can be intercepted, notably when an eavesdropper is placed close to the legitimate transmitter \cite{LopezMartinexPJ15}. In practice, this requires some intruder prevention practices, such as installing a shield behind the FSO receiver to block the undetected light in situations when the incident beam area is larger than the receiving aperture, as proposed in \cite{fsonaSecrecy} (for a commercial FSO product).

\subsection{Channel Propagation Effects}
\label{subsec:ChannelEffects}
\begin{figure}[b]
        \centering
        \includegraphics[width=\linewidth]{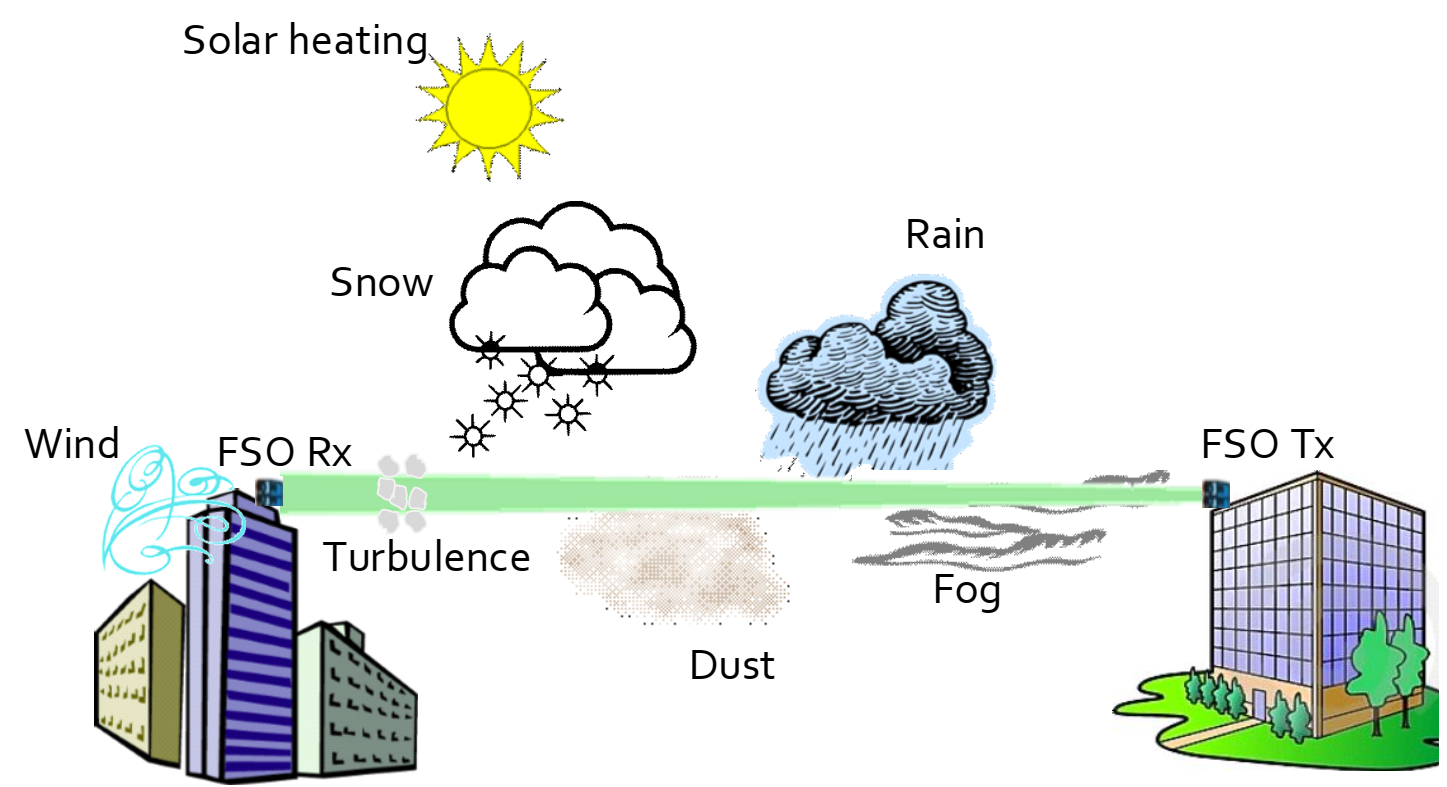}
        \caption{Possible propagation effects on FSO signals transmitted through the atmosphere.}
        \label{fig:environment}
    \end{figure}

\begin{figure}[H]
        \centering
        \includegraphics[width=\linewidth]{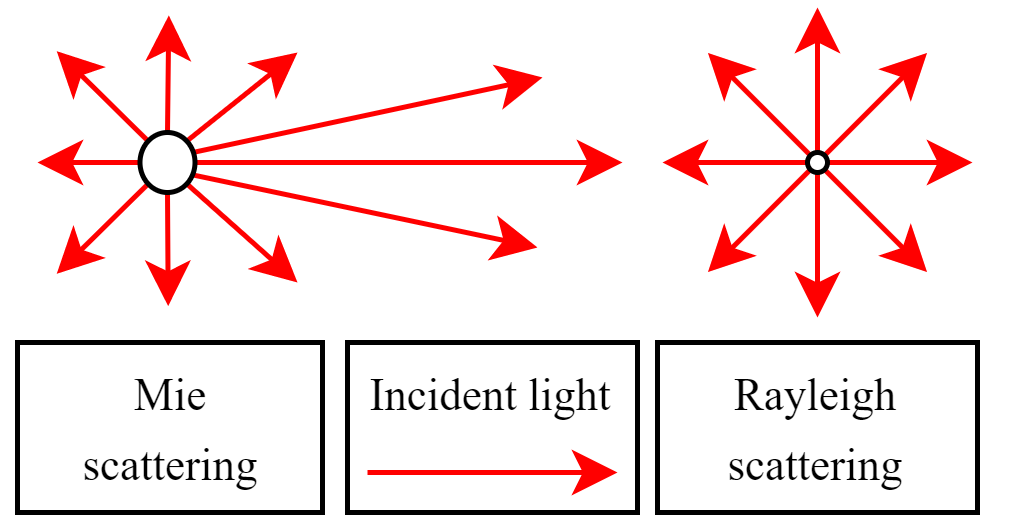}
        \caption{An illustration of Rayleigh and Mie scattering types. Rayleigh is an all-direction scattering, while Mie scattering tends to be in the forward direction.}
        \label{fig:ScatteringTypes}
    \end{figure}

Even though the fundamental principle of both FSO and RF technologies is based on the transmission of electromagnetic waves through the atmosphere, the wavelength and the wave interaction with the outdoor environment vary drastically between the two technologies. Overall, FSO is more sensitive to propagation effects, as will be discussed in this section. The possible propagation effects that may encounter FSO signals are illustrated in Fig. \ref{fig:environment}.\newline
While propagating through the atmosphere, optical signals are subject to attenuation. Attenuation originates from two main effects; absorption and scattering. Absorption is caused by the interaction of waves with molecular and aerosol particles in the atmosphere. However, IR FSO wavelengths are chosen to match atmospheric transmission windows where the molecular and aerosol absorption is minimum.
Therefore, scattering, experienced by IR light waves in the atmosphere, is the major factor for attenuating the received signal. There are two main scattering types when propagating through the atmosphere; Rayleigh scattering \cite{Rayleighscattering} and Mie scattering. As illustrated in Fig.~\ref{fig:ScatteringTypes}, Rayleigh scattering is a diffused all-directions scattering, while Mie scattering is a forward scattering.
Rayleigh scattering occurs when the particles in the atmosphere are much smaller than the wavelength.
Mie scattering is, conversely, the scattering of an EM wave on particles of comparable or larger size than the wavelength. RF waves with wavelengths of $\sim 10^{-2}$ m and larger experiences almost exclusively Rayleigh scattering, whereas FSO with typical wavelengths of $\sim 10^{-6}$ m experiences mainly Mie scattering. Fog is the primary source of Mie scattering. Other weather conditions such as rain, snow, and dust cause a third type of scattering for FSO signals with particle sizes larger than the optical wavelength. \newline 
The effect of fog on FSO systems is significantly detrimental, and the attenuation could reach 200 dB/km and even 350 dB/km in low visibility regimes \cite{7515199}. Attenuation under dusty channels at low visibility could be higher than fog as demonstrated in \cite{7501899} using an FSO experiment with a laboratory-emulated dust storm.  Attenuation due to snow can also reach 150 dB/km \cite{KimSpie11}. Compared to fog and snow, the effect of attenuation is less drastic during the rain, with a 45 dB/km attenuation loss recorded for 1550 nm signals under a heavy rain regime \cite{KimSpie11}. In this view, the combination of FSO and RF in one communication system is even more appealing, as the RF is affected the worst by the rain of all atmospheric conditions, whereas the fog has little to no effect on the RF propagation. \newline

\begin{table*}[t!]
\centering
\caption{\label{tab:Comparison} Comparison between FSO and RF technologies.}
\begin{tabular}{|c|c|c|}
\hline
Technology&RF&FSO\\
\hline
Available Spectrum&$\sim$ 300 GHz&$\sim$ 100s THz\\
\hline
Licensing&Licensed&Unlicensed\\
\hline
Terminal Foot-prints&Bulk&Small\\
\hline
Energy Consumption&Tens of [W]&Tens of [mW]\\
\hline
Data Rates&Mbits/Gbits&Tbits\\
\hline
Configuration&LoS/NLoS&LoS\\
\hline
Latency&Moderate&Low\\
\hline
Noise sources&Interference with other users&Background ambient light\\
\hline
Safety&Subject to EMF radiations&Eye and skin safe\\
\hline
Security&Susceptible to Eavesdropping&Secure\\
\hline
Sensitivity to Pointing Errors&Low&High\\
\hline
Propagation Effects&Attenuation, fading&Attenuation, turbulence\\
\hline
\end{tabular}
\end{table*}
\indent FSO channels are also subject to turbulence resulting from random variations of the refractive index due to fluctuations of temperature and pressure in the atmosphere. Atmospheric turbulence affects FSO almost exclusively due to the wavelength compared to RF. Atmospheric turbulence can be modeled using numerical models based on Kolomogrov theory and statistical models \cite{AndrewsBook}. Different statistical models have been widely used to model turbulent FSO channels in the literature, such as the negative exponential, log-normal, Gamma-Gamma, and the Malaga models \cite{UysalComst14}.\newline
Dynamic wind loads causing sway of buildings, weak earthquakes, and small vibrations due to road traffic cause the beam to deviate from its path randomly \cite{PointingErrors}. These effects are known as pointing errors and impose additional challenges for FSO links and severally affect the transmissions' reliability if not compensated. Pointing errors are commonly modeled using a Rayleigh distribution \cite{PointingModel}. In terms of noise, FSO is subject to ambient noise from the sun, while RF is subject to thermal noise. \newline 
For a full budget link calculation of FSO links, taking into account attenuation, turbulence, beam divergence, and pointing errors, we direct the reader to the detailed study in \cite{TrichiliJOSAB20}.
As takeaway messages of this section, the full comparison between the FSO and RF technologies is summarized in Table~\ref{tab:Comparison}. Moreover, the most common line of sight (LoS) models for the FSO and RF channels under different channel conditions are reported in Table~\ref{tab:models}.

\begin{table*}[]
\centering
\caption{\label{tab:models} Most common LoS models for FSO and RF channels.}
\begin{tabular}{|l|l|l|l|}
\hline
&Atmospheric Condition & Model & Description and Parameters \\
\hline
  \multirow{15}{*}{FSO} &Fog&  Mie scattering model: $\displaystyle A=\frac{13}{V}\left( \frac{\lambda}{0.55} \right)^{-q}$ [dB/km]& \makecell[l]{$\cdot$ Empirical model for scattering due to fog.\\ $\cdot$ Model parameters \cite{kim}: \\$V$ [km]: Visibility range\\ $\lambda$ [$\mu$m]: Operating wavelength\\ $q$: Size distribution of the scattering\\ particles } \\ \cline{2-4}
  &Dust storm & Dust attenuation model: $\displaystyle A_{d}=-52V^{-1.05}$  [dB/km]  & \makecell[l]{$\cdot$ Empirical model for dust scattering.\\$\cdot$ Model parameters \cite{mageddust}:\\$V$ [km]: visibility range\\ $\lambda$=1550 nm } \\\cline{2-4}
  &Snow  & Snow attenuation model:  $\displaystyle A_{s}=k/V$  [dB/km] & \makecell[l]{$\cdot$ Empirical model for snow.\\$\cdot$ Model parameters:\\$V$ [km]: Visibility range\\ $k$: Constant defined in \cite{snow} }\\\cline{2-4}
  &Rain& \makecell[l]{Rain attenuation model: $\displaystyle A_{r}=KR^\alpha$  [dB/km]}& \makecell[l]{$\cdot$ Model describing scattering due to rain.\\$\cdot$ Empirical model for rain scattering.\\$\cdot$ Model parameters:\\$R$ [mm/hr]: Precipitation intensity \\$(K, \alpha)$: Model parameters  \cite{rain}}\\ \cline{2-4} 
    & \makecell[l]{Weak turbulence} &\makecell[l]{Log-normal turbulence model:\\$\displaystyle f(I)=\frac{1}{2I\sqrt{2\pi\sigma_I^2}}\exp\left(-\left(\ln\frac{I}{I_o}+\frac{\sigma_I^2}{2}\right)^2\bigg/2\sigma_I^2\right)$}& \makecell[l]{$\cdot$ Statistical model suitable for weak \\fluctuations regime.\\$\cdot$ Model parameters:\\$E[I]=I_{0}$: Mean irradiance.\\$\sigma_I^2$: Scintillation index (SI) \cite{lognormal}}\\\cline{2-4} 
  & \makecell[l]{Strong turbulence} &\makecell[l]{Negative exponential model:\\$\displaystyle f(I)=\frac{1}{I_{0}}\exp\left(-\frac{I}{I_{0}}\right),~I_{0}>0$}& \makecell[l]{$\cdot$ Statistical model suitable for strong\\ fluctuations regime.\\ $\cdot$ Model parameters:\\ $E[I]=I_{0}$: Mean irradiance}\\\cline{2-4} 
  &\makecell[l]{Moderate to strong\\ turbulence} & \makecell[l]{Gamma-Gamma turbulence model:\\$\displaystyle f_{I}(I)=\frac{2(\alpha\beta)^{(\alpha+\beta)/2}}{\Gamma(\alpha)\Gamma(\beta)}I^{(\alpha+\beta)/2-1}K_{\alpha-\beta}\left(2\sqrt{\alpha\beta}I\right)$} & \makecell[l]{$\cdot$ Statistical model suitable for moderate \\to strong turbulence conditions. \\$\cdot$ Model parameters:\\ $(\alpha,\beta)$: Turbulence related parameters \\$\cdot$$\Gamma(.)$: Gamma function\\ $\cdot$$K_{v}(.)$: Modified Bessel function of the\\ second kind and order $v$} \\ \cline{2-4}
  &\makecell[l]{All turbulence conditions}  &  \makecell[l]{Málaga turbulence model:\\$\displaystyle f_{I}(I)=A\sum_{k=1}^{\beta}\alpha_{k}I^{\frac{\alpha+k}{2}-1}K_{\alpha-k}\left(2\sqrt{\frac{\alpha\beta I}{\gamma\beta+\Omega}}\right)$\\$\displaystyle\begin{cases}
A=\frac{2\alpha^{\alpha/2}}{\gamma^{1+\alpha/2}\Gamma(\alpha)}\left(\frac{\gamma\beta}{\gamma\beta+\Omega}\right)^{\beta+\frac{\alpha}{2}}\\
  \alpha_{k}=\left(\begin{array}{c}\beta-1\\k-1\\\end{array}\right)\frac{(\gamma\beta+\Omega^{'})^{1-\frac{k}{2}}}{(k-1)!}\left(\frac{\Omega^{'}}{\gamma}\right)^{k-1}\left(\frac{\alpha}{\beta}\right)^{\frac{k}{2}} 
\end{cases}$} &\makecell[l]{$\cdot$ Statistical model suitable for all turbulence\\ regimes and can be simplified to Log-normal,\\ negative exponential, and Gamma-Gamma\\ models\\
$\cdot$ Model parameters \cite{MalagaDistribution}\\$\beta$: Amount of fading parameters\\$\Omega:$ Average power of the LoS term\\}\\ 
  \hline
  \multirow{4}{*}{RF} & Fading & \makecell[l]{Nakagami-\textit{m} model\\ $\displaystyle f(x)=\frac{2m^m x^{2m-1}}{\Omega^m\Gamma(m)} \mbox{exp}\left(-\frac{mx^2}{\Omega}\right)$}&\makecell[l]{\\ $\cdot$Suitable to model a wide range of fading.\\ $\cdot$ Model parameters:\\$x$: Channel fading amplitude, \\$m\ge\frac{1}{2}$: Shape factor (fading parameter)\\$\Omega>0$: Mean-square value of $x$ \cite{slimbook}\\$\cdot$$\Gamma(.)$: Gamma function} 
  \\\cline{2-4}
  &Multipath fading&\makecell[l]{Rician model:\\$\displaystyle f(x)=\frac{2(K+1)x}{\Omega}\exp\left(-K-\frac{(K+1)}{x^{2}}\right)I_{0}\left(2\sqrt{\frac{K(K+1)}{\Omega}}x\right)$}& \makecell[l]{$\cdot$  Stochastic model to model multipath \\ propagation with the line of sight path being\\ stronger than others.\\$\cdot$ Model parameters:\\$K$ direct path to other scattered paths power\\ ratio \cite{RiceFading}\\$\cdot$$I_{0}(.)$: Modified Bessel function of the\\ first kind and order 0}\\\cline{2-4}
  \hline

\end{tabular}
\end{table*}

\section{Mixed RF/FSO Multihop Links}
\label{sec:AsymmetricRFFSO}
\begin{figure}[tb]
        \centering
        \includegraphics[width=\linewidth]{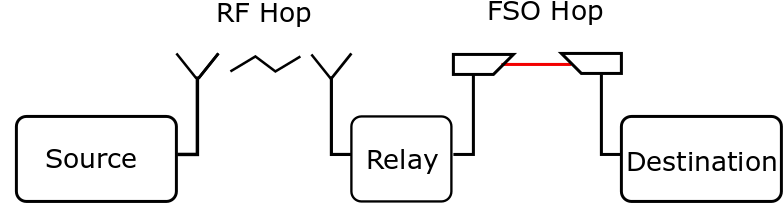}
        \caption{A schematic illustration mixed serial dual-hop RF/FSO configuration. RF signals at the relay node are converted to optical signals and forwarded to the destination on the FSO link.}
        \label{fig:DualHopRFFSO}
\end{figure} 
A possible scenario for the deployment of FSO is when it is combined with RF transmission systems to fill the connectivity gap between the RF access network and the fiber backbone network \cite{RFFSOLee, RFFSOSamimi}. A schematic illustration of a dual-hop mixed RF/FSO configuration is shown in Fig.~\ref{fig:DualHopRFFSO}. The dual-hop has two primary designs: Amplify and Forward (AF) and Decode and Forward (DF).\newline
\begin{itemize}
    \item In an AF system, the data initially encoded on an RF channel is received by an antenna at a relay node. The received electrical signals are converted to optical signals and amplified and forwarded to the FSO transmitter at the relay.
    \item In the case of DF relaying, the signal is decoded, regenerated, and then retransmitted to the FSO node.  \newline
\end{itemize}
The AF approach is preferable and less complex than DF, as it does not require any decoding at the relay node.\newline
\indent Dual-hop heterogeneous systems, with both AF and DF relays, have been well studied considering a wide range of RF and FSO channels \cite{RFFSOLee, RFFSOSamimi,ZediniPJ15, AneesIET15, DualHopGeneralizedRFFSO}. In a seminal study, Lee \textit{et al.} studied the outage performance of a dual-hop heterogeneous RF/FSO system with a fixed gain AF relay \cite{RFFSOLee}. The RF link is assumed to be subject to a Rayleigh fading, and the FSO is subject to Gamma-Gamma turbulence. A sub-intensity modulation (SIM) scheme was considered for the AF relay to convert the incoming RF signal to optical signals before being forwarded to the destination. We note that SIM is a technique that consists of modulating a pre-modulated RF signal on the intensity of optical signals, which lowers the system's complexity \cite{SIMModulation}. The primary outcome of the study reported in \cite{RFFSOLee} is that RF/FSO systems have slightly worse performance  (in terms of outage probability) than conventional dual-hop RF/RF links. The work was followed by the outage performance analysis of a dual-hop system subject to a Rayleigh fading for the RF hop and an M-fading generalized model for the FSO hop \cite{RFFSOSamimi}. The authors also assumed a SIM scheme to convert the RF signals to optical signals at the fixed-gain AF relay node and further considered the effect of pointing errors that could occur on the FSO link. The capacity and error performances of a dual-hop RF/FSO system with a Nakagami-$m$ RF channel and a Gamma-Gamma distributed FSO link were studied in \cite{ZediniPJ15}. Authors of \cite{AneesIET15} considered the same RF and FSO channel conditions as \cite{RFFSOLee} but with a DF relaying scheme. Both fixed gain and channel state information (CSI) relaying scenarios are considered in a dual-hop RF/FSO study reported in \cite{DualHopGeneralizedRFFSO}, where the RF link is subject to a Nakagami-$m$ fading, and the FSO is subject to Gamma-Gamma turbulence. Note that the CSI-based relaying, also known as variable gain, assumes feedback sent from the receiver to the transmitter. The effect of outdated CSI on the outage BER performances of RF/FSO mixed system was investigated in \cite{OutdatedCSI}. The authors assumed that the RF link is subject to Rayleigh fading, and the FSO is modeled by a Gamma-Gamma distribution.  \newline
\indent A less conventional relaying scheme known as Quantize and Encode (QE) was considered in an RF/FSO dual-hop system \cite{QuantizeEncode}. In a QE relaying scheme, the RF source uses a quadrature amplitude modulation (QAM) format, and the relay estimates and quantizes the log-likelihood ratio of each received bit in a QAM symbol and then transmits it through the FSO channel. 
Overall, with many relaying schemes, mixed RF/FSO systems have been well studied in the literature.

\section{Hybrid RF/FSO system: Operation and Switching Mechanisms}
\label{sec:HybridRFFSO}
\begin{figure}[tb]
        \centering
        \includegraphics[width=\linewidth]{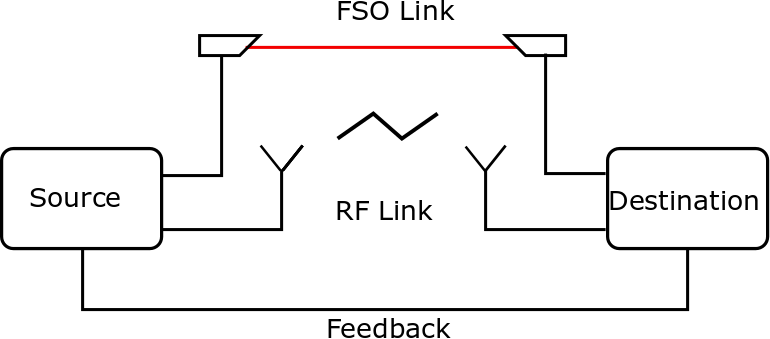}
        \caption{A schematic illustration of a hybrid RF/FSO configuration. Data can be transferred on the RF or the FSO link, or both.}
        \label{fig:HybridRFFSO}
\end{figure}
Besides being used in a dual-hop fashion, as discussed in section \ref{sec:AsymmetricRFFSO}, RF/FSO technologies can be installed in parallel, forming a hybrid communication system. An illustration of a hybrid RF/FSO system that supports the simultaneous transmission of data using RF and FSO signals is depicted in Fig.~\ref{fig:HybridRFFSO}. 
This section discusses the various operation modes of hybrid RF/FSO systems from what has been proposed in the literature. We start by presenting two switching mechanisms between the RF and FSO links, known as the hard and soft switching schemes. A third scheme based on the use of machine learning algorithms is then presented.  We further cover the case when both the RF and FSO transmitters operate simultaneously, known as the diversity scheme. 

\subsection{Hard Switching}

\begin{figure}[H]
        \centering
        \includegraphics[width=\linewidth]{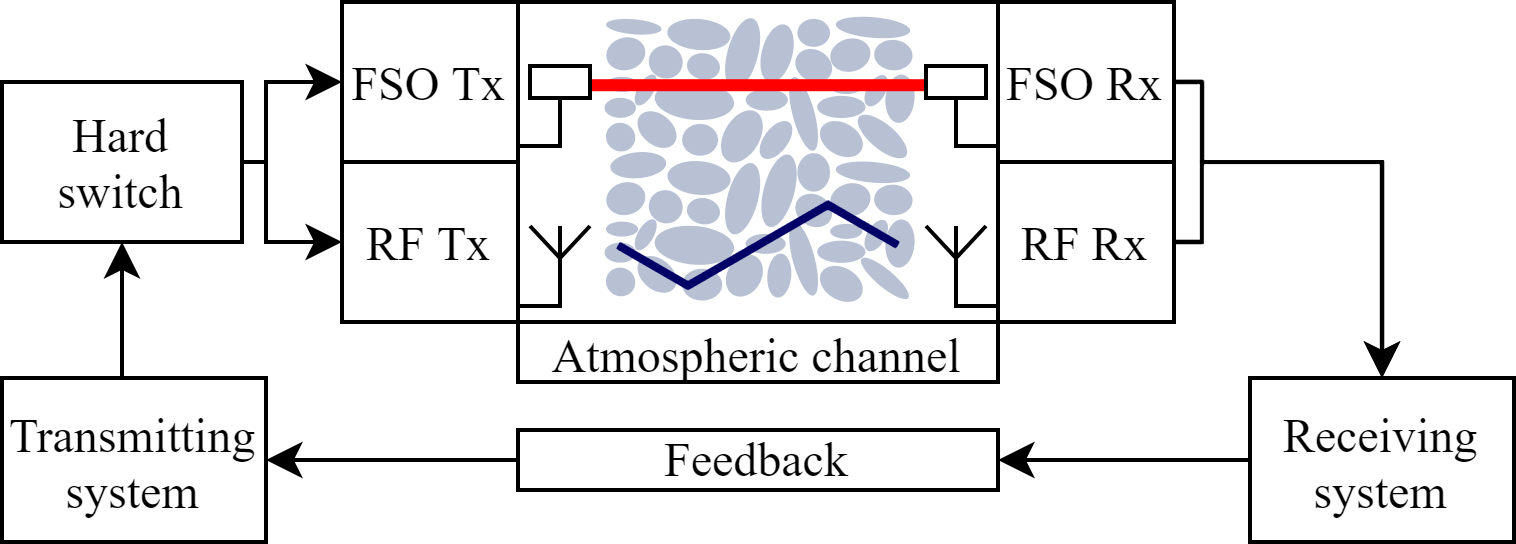}
        \caption{An illustration of a hard switching system diagram. Continuous feedback from the receiver to the transmitter is needed.}
        \label{fig:Hard_switch}
\end{figure}
In a system based on hard switching, either the RF link or the FSO link is operational at any given time. Feedback from the receiver to the transmitter is used to coordinate between the two communicating terminals to perform the switching. \newline
Usman \textit{et al.} proposed in \cite{UsmanPJ14} a hard switching mechanism for hybrid FSO/RF systems. The authors assumed the use of the FSO link when it is above a certain threshold, and when the link becomes unacceptable, the RF is used instead. A dual FSO threshold system design was also considered, where the FSO has two different thresholds for transitioning between ``on'' and ``off'' states. The dual threshold system showed to be similar in performance to a simple hard switching while reducing the number of switches and preserving the FSO subsystem. The performance of the hard-switching approach was further investigated in numerous manuscripts \cite{VISHWAKARMA2021126796,AbadiICC17}. Authors of \cite{VISHWAKARMA2021126796} studied the performance of a hard-switch-based RF/FSO system over generalized fading models. Time Hysteresis (TH) and Power Hysteresis (PH) modifications to the hard-switching approach were studied in \cite{AbadiICC17}, where the switching threshold may be variable based on the channel conditions and history. \newline 
A multi-user FSO/RF hybrid network scheme was proposed in \cite{RakiaTWC20}, where a primary FSO link services every user. When the FSO link fails, a central node allocates a backup RF link with non-equal priority to different users. Finally, the performance of a selective dual-hop RF/FSO DF relay network based on hard switching was studied in \cite{Sharma:19}. The authors assumed two RF/FSO subsystems with a relay node and maximum ratio combining (MRC) scheme at the receiver \cite{Sharma:19}. The proposed scheme was found to outperform FSO cooperative system and single hope RF/FSO system. 
\subsection{Soft Switching}
\begin{figure}[H]
        \centering
        \includegraphics[width=\linewidth]{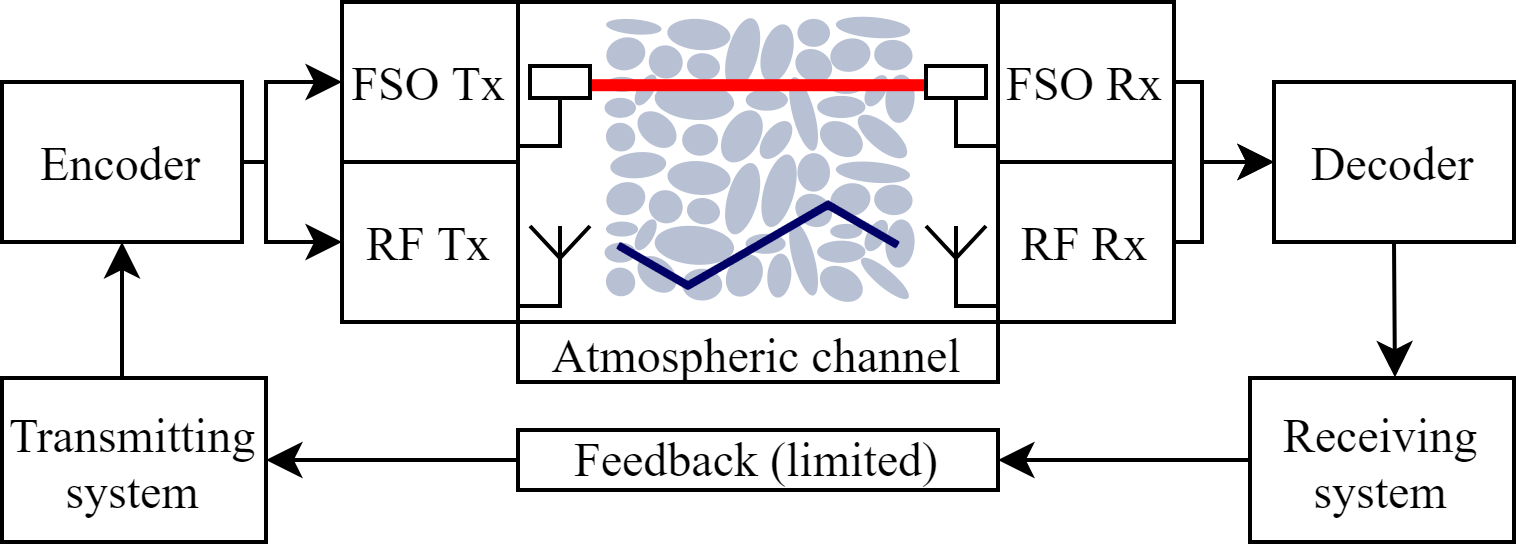}
        \caption{An illustration of a soft switching system diagram based on joint encoding and decoding and with an optional (that is limited if used) feedback from the receiver to the transmitter.}
        \label{fig:Soft_switch}
\end{figure}
Within the soft-switching approach, channel coding can be used to coordinate between the FSO and the RF links. The main advantage of the soft-switching compared to the hard-switching is that the rate of the FSO link is not wasted if the RF link is selected. Zhang \textit{et al.} proposed a soft switching approach based on short length Raptor codes \cite{ZhangJSAC09}. In Raptor codes, each message sent from the RF and FSO subsystems is different from the previous one. The decoder decides to accept the entire block as soon as a sufficient number of messages have been collected, whether they come from RF or the FSO transmitter. This approach allows utilizing an otherwise unused FSO data rate for long-distance transmission scenarios when the received FSO power would be considered below the threshold for the hard switching-based mechanism. Authors of \cite{ZhangJSAC09} showed a rate gain by 4 folds (and more) compared to the hard switching approach for  1 and 2 km transmission length scenarios. However, both hard- and soft switching approaches showed similar average data rate performance at a half-km transmission length due to the sufficient optical power at the receiver. \\
Alternatively, authors of \cite{TangTCOM12} proposed to adapt the transmitted symbol rate and the constellations on the RF and FSO links depending on the channel conditions. The authors pointed that such an approach could be particularly of interest for grid-connected systems, where power conservation is not crucial.

\subsection{Machine Learning Based Switching}
\begin{figure}[H]
        \centering
        \includegraphics[width=\linewidth]{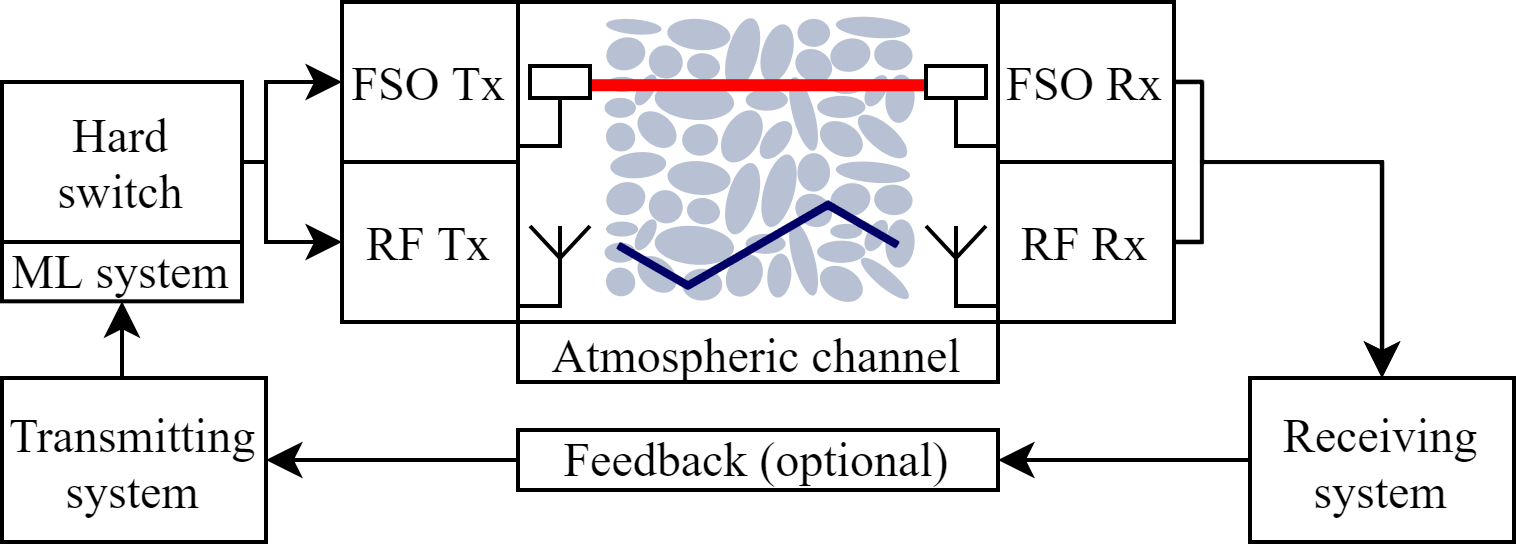}
        \caption{An illustration of an ML-powered system diagram where hard switching is performed based on the ML prediction. The feedback in the ML scheme is optional.}
        \label{fig:Diversity}
\end{figure}
Machine learning is a versatile tool that has found uses in almost all the various fields of communication. Machine learning has been used in RF and optical communication systems for applications such as performance monitoring, modulation format identification, channel modeling, and digital signal processing \cite{9184046,8527529,8666641}. However, for the hybrid FSO/RF systems, only a few studies have been so far reported. In \cite{haluvska2020} and \cite{toth2018c}, several machine learning algorithms, including random forest, decision tree, and others, are exploited to predict the received signal strength indicator (RSSI) in a hybrid FSO/RF hard switching system. Accurate prediction of RSSI parameter helps in fastening the switching between FSO link and RF link where for low RSSI, low-speed RF link is used and for high RSSI, high-speed FSO link is used. The obtained results show high accuracy for predicting the RSSI parameter with superiority for the decision tree algorithm.
Proactive switching for hybrid FSO/RF links that ensure energy efficiency is studied in \cite{meng2019}. In hybrid FSO/RF links, energy wastage happens because of the need to continuously keep the FSO link ``on'' when the RF link is operational. The reason is to monitor the channel, so when the channel conditions improve, the system switches to the FSO link. To cope with this problem, the authors of \cite{meng2019} proposed that when the RF link is ``on'', the FSO link is kept in sleep mode with the capability to activate it periodically to sample the FSO signal power by transmitting beacon signals. These samples are used to train a Long Short Term Memory (LSTM) algorithm to predict the link switching time (the time when the FSO signal is expected to be above a predetermined threshold). The reported simulation results have shown that the proposed technique can significantly improve the energy efficiency of the RF/FSO system \cite{meng2019}.
\subsection{Diversity Scheme}
\begin{figure}[H]
        \centering
        \includegraphics[width=\linewidth]{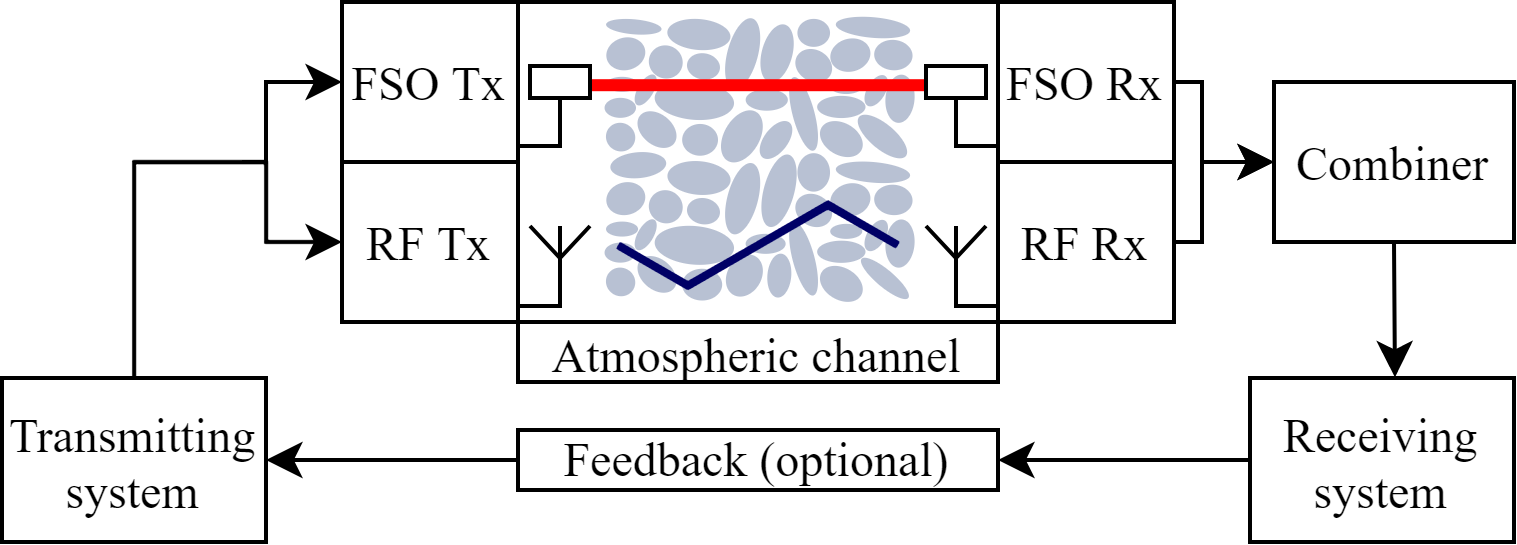}
        \caption{An illustration of a diversity system diagram where both systems simultaneously and the received data from the RF and FSO link are combined in a combiner unit. The feedback in the diversity scheme is optional.}
        \label{fig:Diversity}
\end{figure}
The diversity scheme consists of simultaneously transmitting the same information on FSO and RF co-existing links.
Under the assumption that millimeter waves at 6 GHz support the same data rates as optical links, authors of \cite{FSORFDiversity} proposed encoding the same data signals on an FSO and an RF link. At the receiver, selective combining (SC) and MRC schemes are considered. This diversity approach is challenging mainly when a complex modulation format, such as phase shift keying (PSK), is used with the optical links requiring coherent systems.\newline 
\indent Authors of \cite{AMIRABADI2019293} proposed the use of receive diversity in a hybrid setup by introducing multiple receivers for both FSO and RF subsystems. The FSO link is assumed to have an equal gain combining (EGC) scheme, while the RF had an MRC scheme to collect received signals. An IM/DD configuration was considered for the FSO, making the implementation of such a system possible in practice. The proposed hybrid RF/FSO system design demonstrated reduced power consumption for a similar performance compared with RF or FSO only solutions and the case when only a single receiver RF/FSO system is used. The primary motivation of the proposed technique is to be used in scenarios where one of two links experiences a frequent outage due to the weather conditions.\newline 
Rakia \textit{et al.} proposed a power adaptation scheme based on truncated channel inversion for an FSO/RF system employing adaptive combining to ensure a constant signal-to-noise ratio (SNR) \cite{RakiaPJ15}.
The FSO transmitting power was considered to be constant, and two policies of power allocation to the RF subsystem, based on RF only and joined SNR, were considered. Unlike other diversity schemes presented, the feedback channel between receiver and transmitter is once again needed in this scheme. \newline
\indent Finally, a complex hybrid RF/FSO backhaul network was considered in \cite{8883295}. The authors proposed the use of buffer-aided and quality-of-service (QoS) aware relays in conjunction with novel adaptive single-carrier and multiple carrier transmission schemes to reduce delay for latency-sensitive applications such as virtual reality and smart grid control. Given a fixed QoS, numerical simulations showed an increase in the maximum supported arrival rate.\newline 
A detailed summary of the studies on the various RF/FSO switching mechanisms in the literature is given in Table~\ref{Tab:SwitchingSummary}.

\begin{table*}[h!]
    \centering
     \caption{Summary of switching mechanisms for hybrid RF/FSO systems studied and demonstrated in the literature. } 
     \begin{tabular}{|m{20pt}|m{50pt}|m{100pt}|m{60pt}|m{77pt}|m{120pt}|}
     \hline 
     Ref.  & Study Type & Scope & Hardware Configuration & Investigated Metrics & Main Conclusions \\\hline 
     
     \cite{UsmanPJ14} & Analytical, with numerical examples & Hard switching with single and double threshold policy & RF/FSO switching with feedback control & -- Outage probability\newline -- Average BER for non-outage time periods\newline -- Ergodic capacity & The dual threshold achieves performance similar to that of single threshold policy but increases the expected lifetime of the FSO link.  \\\hline 
     
     \cite{VISHWAKARMA2021126796} & Analytical / numerical & Hard switching with single threshold policy over generalized fading models & RF/FSO switching with feedback control & -- Outage probability \newline -- Average symbol error rate (SER) \newline -- Ergodic capacity & The range of optimum FSO beam waists and optimum switching thresholds are provided. RF/FSO links are found to perform better than FSO only systems under strong atmospheric turbulence and high pointing errors.  \\\hline 
    
    \cite{AbadiICC17} & Algorithmic/ numerical & Hard switching with TH and PH switching schemes & RF/FSO switching with feedback control & -- Average hybrid data rate \newline -- Link availability & The high data rate of individual links does not directly correlate with the high data rate of the hybrid link. \\\hline 
     
     \cite{Sharma:19} & Analytical, with numerical examples & Hard switching with a single threshold for selective DF relay network & RF/FSO switching with feedback control & -- Outage probability \newline -- Average SER & The proposed system increases the performance when compared to the single-hop hybrid system and cooperative FSO system with MRC combining scheme. \\\hline

     \cite{RakiaTWC20} & Analytical, with numerical examples & Hard switching with a single threshold for multi-user network scenario under non-equal priority & RF/FSO switching with feedback control & -- Average buffer size\newline-- Average buffer queuing delay\newline-- Frame loss probability\newline -- Efficiency of the queue \newline -- RF link utilization \newline -- Throughput  & The proposed switching algorithm ensures higher performance for prioritized nodes, while still serving the low-priority nodes. The common backup RF link also improves the performance of all the nodes in the network. \\\hline 
     
     \cite{ZhangJSAC09} & Algorithmic / experimental / numerical & Soft-switching with Raptor codes & RF/FSO joint encoding and decoding & -- Daily average data rate\newline Data rate probability densities for an experimental scenario based on meteorological data & Short-length Raptor code based encoders nearly quadruples the average data rate on the studied dataset, compared to the hard switching. \\\hline 
     
     \cite{TangTCOM12} & Analytical with numerical examples & Soft-switching with link adaptation & RF/FSO joint encoding and decoding & -- Throughput in relation to channel conditions & A gain of about 10\% on the considered dataset is reported, when compared to fixed a modulation scenario. \\\hline 
     
     \cite{haluvska2020} & Algorithmic / experimental / numerical & RSSI for ML-powered hard switching & RF/FSO switching based on RSSI & -- Comparison of 3 ML models performance for RSSI prediction. & The appropriate values for the ML model parameters were studied. \\\hline
     
    \cite{toth2018c} & Algorithmic / experimental / numerical & RSSI for ML-powered hard switching  & RF/FSO switching based on RSSI & --Comparison of 3 ML models performance for the RSSI prediction in binary and multi-class implementation. & Accurate prediction of RSSI can enhance the efficiency of hard FSO/RF switching. \\\hline
     
     \cite{meng2019} & Algorithmic / experimental / numerical & Predictive link switching & RF/FSO predictive hard switching & -- Energy consumption\newline -- Received optical signal power\newline -- Energy efficiency &The predictive switching improved the energy efficiency of the system compared to the reactive link switching
     \\\hline 
     
     \cite{FSORFDiversity} & Analytical, with numerical examples & Transmit diversity with the SC and MRC combining schemes for PSK modulation & RF/FSO joint decoding with no receiver feedback & -- BER performance & MRC is the most optimal combining method for the proposed RF/FSO scenario. \\\hline 
     
     \cite{AMIRABADI2019293} & Analytical, with numerical examples & Receive diversity with EGC scheme for the FSO link and MRC scheme for the RF & SC-based RF/FSO receiver with no feedback &-- Outage probability & The proposed system is recommended for use in Mediterranean climate due to heavy rain/fog.\\\hline 
     
     \cite{RakiaPJ15} & Analytical with numerical examples & Adaptive RF switching and power adaptation based on FSO link quality with MRC receiver & RF/FSO MSC with limited feedback for RF on/off & -- Outage probability\newline -- Outage capacity & The system has shown improvement in performance in terms of outage probability and RF subsystem power conservation compared to baseline.\\\hline 
     
     \cite{8883295} & Analytical with numerical examples & Adaptive transmission for single-carrier, and multiple-carrier hybrid RF/FSO with relaying  & Receiver only with single-carrier or multiple-carrier with no feedback &-- Maximum supported arrival rate\newline -- Average queuing-delay\newline -- RF and FSO data rate contribution & The system has shown to improve the QoS aware maximum supportable arrival rate of the hybrid RF/FSO network.\\\hline 
     \end{tabular}
     \label{Tab:SwitchingSummary}
\end{table*}

\begin{table*}[t!]
    \centering
     \caption{Configuration of the current commercial hybrid FSO/RF systems}
    \begin{tabular}{|m{60pt}|m{45pt}|m{55pt}|m{30pt}|m{30pt}|m{30pt}|m{40pt}|m{30pt}|m{30pt}|m{25pt}|m{10pt}|}
    \hline
        \multirow{2}{*}{Company name \vspace{-10 pt}} & \multirow{2}{*}{Product name \vspace{-10 pt}} & \multirow{2}{*}{Country of origin \vspace{-10 pt}} & \multicolumn{2}{c|}{Data rate [Gbps]} & \multirow{2}{*}{Reach\vspace{4 pt}} & \multirow{2}{*}{Laser \vspace{4 pt} } & \multirow{2}{*}{RF band\vspace{-
        3 pt}} & \multirow{2}{*}{Latency} &
        \multirow{2}{*}{Ap. cost\vspace{-3 pt}}&
        \multirow{2}{*}{Ref.\vspace{-10 pt}}\\
        \cline{4-5}
        & & & FSO & RF & distance [km] & wavelength [nm] & (GHz) & (msec) & (k~USD)& \\\hline 
        
        Communication by light (CBL) & AirLaser IP1000plus & Germany & 1 & 0.1 & Up to 1 & 830 -- 870& 5 &0.052 & $\sim$18 & \cite{cbl_datasheet} \\\hline
         JV-Labs.EU & iRedStar & Czech Republic & 1 & 0.8 & 0.7 & 850 & 5 & 0.5 & $\sim$5  &\cite{jv-labs}\\\hline
          \multirow{2}{*}{LightPointe\vspace{-12pt}} & FlightStrata 100 XA & \multirow{2}{*}{USA\vspace{-12pt}} & 0.1 & 0.072 & 5 & 850 & 5.8 &$\leq$0.00002&NA  &\cite{teldata}\\
           \cline{2-2} 
           \cline{4-11}
           & HyBridge SXR-5& & 1 & 0.15 & 5 & 850 & 5.4/5.8 &$\leq$0.00002 & $\sim$16 &\cite{lightpointe}\\
            \hline
           Mostcom (ARTOLINK) & Artolink M1-30GE & Russia & Up to 30 &N/A & 1.3 & 1550 & E-band & 0.005  &10-15 & \cite{MostCom}\\\hline
           Collinear & HybridCNX-520  & Australia  & 10 & Up to 9.8 & Up to 5 & 1550 & \makecell[l]{71-76/\\81-86} &ULow&NA &\cite{collinear}\\\hline
            EC Systems & EL-10Gex & Czech Republic & Up to 30 &N/A & 1.3 & 1550 &E-band & $\leq$0.005 &NA &\cite{ECsystems}\\\hline
            Trimble & PXW3500GT & Hungary & 1 &N/A & 3.5 & 785 & WiFi & 0.00005  &$\sim$10&\cite{Trimble}\\\hline
              fSONA & SONAbeam 10G-E+ & Canada & 10 & 0.1 & Up to 5 & 1550& WiFi & Low &$\sim$25 &\cite{sona}\\\hline
              Hyperion Technologies & Small Cell Backhaul & Turkey & 10 &N/A & 0.3 & 1550 & 2.4 & 0.1  &NA &\cite{hyper}\\\hline
               ANOVA Financial Networks & Celeras LTS & USA & 4 &N/A & 0.3 & 1550 & MMW  & UL  &NA &\cite{anova}\\\hline
             
    \end{tabular}
    \label{tab:commercial}
    \begin{tablenotes}
   \item[*] N/A: Information not available. 
   \item[**] Low/ULow: No precise value was provided for the latency but described as Low/Ultra-Low.
   \item[***] E-band (WiFi): Only information available is that the link operates in the E-band (WiFi).
   \item[****] MMW: Exact MMW frequency/band not provided.
  \end{tablenotes}
\end{table*}
\section{RF/FSO Systems Hardware Configuration}
\label{Sec:Hardware}
In this section, we discuss the hardware configuration for practical hybrid RF/FSO systems. In particular, we review the specification of commercially available RF/FSO solutions. Then, we present some of the recent field trials aiming to set reliable RF/FSO systems.  
\subsection{Commercially Available RF/FSO Solutions}
Recent market statistics have shown that the expected investments in FSO equipment will exceed \$300 million in 2029 \cite{bussinesswire}. There are many commercially available FSO products, and some of these systems offer the possibility to also transmit data over RF channels \cite{MostCom}. In this section, we highlight the main features of some of the currently available FSO/RF solutions. Table \ref{tab:commercial} summarizes the main configurations of commercially available products in terms of data rate, reach distance, operating wavelength, backup RF band, latency, and the approximate cost (whenever the information is available). Note that the system's approximate cost varies with the customization and how big the equipment order is. 

Some hybrid commercial products offer FSO links that can support up to 30 Gbps in ideal weather conditions, such as EC Systems (EL-10Gex) and Mostcom (Artolink M1-30GE). The maximum transmission distance can also reach up to 5 km with Collinear (HybridCNX-520) and fSONA (SONAbeam 10G-E+) solutions. However, in harsh environments (under heavy fog, dust, or snow), the communication link is often switched to the RF backup channel. The data rates are reduced by more than an order of magnitude, as specified by  Communication by Light (CBL-AirLaserIP1000plus), LightPointe (HyBridge SXR-5), and fSONA (SONAbeam 10G-E+) systems. Note that most companies have used unlicensed RF bands for the radio backup links to reduce interference and cost. These bands include both lower frequency sub-bands (typically 2.4 and 5 GHz) and higher frequency millimeter-wave (MMW) sub-bands beyond 60 GHz.

Latency is a crucial parameter in emerging and next-generation wireless networks. For instance, the 5G New Radio (NR) global standard is targeting 1 msec as air-interface latency. In this regard, current commercial products have shown FSO links latency varying from  0.1 msec (Hyperion Technologies) to 50 nsec (Trimble) with transmission distances ranging from 0.3 to 3.5 km, respectively.
This motivates the deployment of these systems in advanced wireless communication networks. Furthermore, the current commercial solutions operate in the 850 nm and 1550 nm bands. Both bands are characterized by low atmospheric attenuation. Besides being ``eye-safe'', one of the advantages of the 1550 nm band is the relatively low cost of FSO transmitting and receiving devices. Advanced vertical-cavity surface-emitting laser (VCSEL) transmitters and high sensitive avalanche photodiodes (APD)  are widely used in this band. The proliferation of high-speed transmitters, amplifiers, wavelength division multiplexing (WDM) components, and receivers optimize the FSO transmission in the 1550 nm band. It is worth noting that the maximum commercial speed in the 850 nm band is 1 Gbps using VCSELs and on-off keying (OOK) modulation, whereas beyond 10 Gbps (OOK) can be transmitted using the enhanced small form-factor pluggable (SFP+) network interface module, 1550 nm.

\subsection{Field Trials of Hybrid RF/FSO Systems}
The outcome of a two-year RF/FSO experiment, conducted between two campuses in the University of Ankara, was reported in \cite{1505831}. For a link distance of 2.9 km, the maximum achieved data rates by the FSO and RF links were 155 and 11 Mbps, respectively. Based on the detected received power of the FSO channel, a hard switching mechanism was set to switch between the two parallel links. Over the experiment period, the RF/FSO links were subject to various natural atmospheric conditions, including snow, haze, cloud, and fog. The author showed snowstorms and fog are the most detrimental conditions that affect the laser link's availability. It is worth noting that the overall FSO link availability exceeded 99.98\% over an observation period. Besides, in \cite{10.1117/12.2546441}, a hybrid outdoor RF/FSO system was tested in Rome, by the Air Force Research Laboratory (AFRL), for a duration of one week. The hybrid system is demonstrated over a 30 km distance between two 100-foot height towers. The data throughput was up to 10 Gbps and 1 Mbps for the FSO and RF channels, respectively. For a reporting duration of 30 minutes, the communication was shifted 5 times ($\sim$ 7 minutes) to the RF link owing to the FSO link outages in foggy weather. Authors of \cite{RFFSODemonstration} demonstrated a hybrid RF/FSO setup that mimics the communication for airborne. The field trial experiment was demonstrated on the German Aerospace Center (DLR) campus near Munich over a 300-meter distance between static objects. The reported data rates were 600 Mbps and 6 Mbps for the FSO and RF link, respectively. A hard switching algorithm is employed such that only one link is used for data transmission if the other is not available. The interruption in the FSO link is demonstrated by applying deterministic fades using an optical attenuator. Another field trial of FSO communication with an RF backup link has been demonstrated in Qatar University \cite{7181911}. Contrary to the other field trials where the effect of fog and snow highly disturb the FSO link, the humidity effect has been considered in this experiment owing to the atmospheric condition in this Gulf area. A communication link of 600 meters was established between two buildings. The authors conducted the outdoor experiment over a period of two months from December to January, where the average humidity and temperature were 66\% and 25 $^{\circ}$C, respectively. The field trial shows the effect of humidity on the FSO throughput performance. The same authors repeated their investigation in \cite{10.1117/12.2256792}. However, the experiments were conducted in the hot summer for the period between June and September, where the average temperature reaches 45 $^{\circ}$C. From both studies, it has been shown that the FSO link throughput is affected more by temperature rather than humidity.

\subsection{Lessons Learned }
In this section, we reviewed the specifications of commercial products and field trials testing hybrid RF/FSO communication systems. The following lessons can be drawn:
\begin{itemize}
    \item Commercial systems with VCSEL-based FSO transceivers are limited to data rates in the level of $\sim$1 Gbps. However, SPF+-based transceivers can provide up to 30 Gbps and enable higher transmission distances compared to VCSEL-based ones.  
    \item Link throughput versus weather conditions was the main focus in most field trials where hard switching mechanism was mainly used, owing to its simplicity. However, considering other switching strategies requires further studies and investigations.
    \item Snowstorm, fog, humidity, and high temperature (i.e., above 40 $^{\circ}$C) are the significant weather conditions that deteriorate the performance of FSO links. However, further field trials under other extreme weather conditions, such as rain and dust storms, are required before large-scale technology deployment.
\end{itemize}

\section{RF/FSO for Space Communications}
\label{Sec:SpaceRFFSO}
Here, we discuss the potentials of the co-existence of RF and FSO technologies for near-Earth and deep space applications.
\subsection{Overview on Satellites and Space Communication}
Satellites have a wide range of applications, including remote sensing, climate change control, deep space exploration, tackling the digital divide, and providing connectivity to rural areas  \cite{SaeedCubesat}.
There are three main types of Earth orbits satellites:\newline
\begin{itemize}
\item Low Earth Orbit (LEO) Satellites: which operate in close proximity of Earth over distances between 160 to 2000 km. LEO satellites are mainly used for imaging and are known for the low latency due to the relatively short propagation distance in space. Recently, LEO satellite constellations (a group of satellites working together) are receiving considerable attention and being used to provide connectivity in rural areas.
\item Medium Earth Orbit (MEO) Satellites: which operate in the region above 2000 km and below the static geostationary orbit. MEO satellites are mainly used for navigation, communication, and space environment science.
\item Geosynchronous Earth Orbit (GEO) satellites: which are fixed with Earth in the geostationary orbit (35,786 km) above the Earth's equator and rotate in the same direction with the Earth's rotation. GEO satellites are typically used for TV broadcasting and used by some satellite communication companies to provide world coverage with a minimum number of satellites.
 \end{itemize}
 In satellite communication, the link connecting the network to the satellite is known as a feeder link. The satellite sends signals to ground users through multi-beams, known as the user links. \newline
\indent Traditional satellite communication has been historically based on microwave bands, in particular, the L (1-2 GHz), Ku-band (12-18 GHz), and the Ka-band (26.4-40 GHz). Compared to the Ku and Ka bands, the L band is less susceptible to fading caused by rain. Ka and Ku have wider bandwidths and smaller components. In the Ku band, only  1 GHz of bandwidth is allocated for satellite services: half of it for uplink and the other half for the downlink. 6 GHz of bandwidth is dedicated in the Ka-band for satellite communication \cite{DLProjects}.

\subsection{Optical Near-Earth Space Communication}
Optical communication in space has made tremendous progress over the years \cite{Hemmati}. An early inter-satellite demonstration has reported a successful error-free 5.6 Gbps transmission between two Low Earth Orbit (LEO) satellites (NFIRE and TerraSar-X) over several seconds \cite{NFIRETSX,NFIRETSXEDRS}. Ground-to-satellite (uplink) and satellite-to-ground (downlink) FSO links have also been demonstrated \cite{ETSVI,Sota}. The first uplink and downlink demonstrations date back to 1994 with a 1.024 Mbps two-way communication between the Japanese Engineering Test Satellite-VI (ETS-VI) and a ground station \cite{ETSVI}. In 2016, the SOTA (Small Optical TrAnsponder) system, developed by the National Institute of Information and Communications Technology (NICT) in Japan, was used to establish an optical communication link between a LEO satellite (SOCRATES) to ground connection with a ground station affiliated to the German aerospace center \cite{Sota}. NASA, in collaboration with MIT Lincoln Laboratory, is currently developing a 200 Gbps optical communication system to be installed on an LEO Cubesat (small satellite at an altitude of 740 km) with a downlink that could reach 200 Gbps to deliver more than 50 Terabytes of information per day to a ground station, as a part of the TeraByte InfraRed Delivery (TBIRD) program \cite{TBIRD}. In addition to the vast bandwidth, optical satellite link require small footprints devices that consume less energy compared to those operating in the microwave bands typically used for Earth-satellite communications.\newline
\indent As the satellite communication race gears up, multiple companies are developing optical free space feeder links and inter-satellite links to bring fiber-like throughput to space. For example, Lightspeed-Telesat, and Starlink-SpaceX projects aim to establish FSO satellite crosslinks in their satellite constellations installed to provide broadband internet access all over the globe \cite{Starlink,Telsat}.\newline
\indent Optical links are preferable compared to RF links for space applications because of the lower divergence of optical waves than the RF ones. However, ground-space optical links are subject to turbulence and weather conditions, particularly in the lower layers of the atmosphere, as discussed in section~\ref{subsec:ChannelEffects}.  \newline
 
\subsection{Mixing Optical and RF Links For Satellite Communication}
Mixing optical communications and RF can be highly advantageous in applications related to satellite, and deep-space communications \cite{Space_FSO}. As we pointed out earlier, optical links are subject to atmospheric turbulence affecting earth-to-space and space-to-earth links. Even though RF and optical beams may propagate through the same geometrical path through the atmosphere, the channel effects, in most cases, are more severe on optical signals. These effects are caused by the various air densities on different heights above the ground and the difference in the air's turbulent behavior closer to the ground. The distortions of the beam attributed to the atmospheric turbulence are far greater close to the ground due to more substantial turbulence and higher air density. This means that the downlink beam will encounter most of the disturbances closer to the receiver, resulting in fewer distortions on the beams' intensity profiles than in the uplink case, where most of the atmospheric disturbances are encountered immediately after leaving the transmitter, and its adverse diffractive effects can accumulate over longer propagation distances. \\ 
Therefore, it may be beneficial for the hybrid RF/FSO system to use FSO for downlink and RF for the uplink. This is also beneficial, as RF downlink signals may contribute to the spectrum congestion on the ground due to its large spot size. We note that RF signals can result in interference for the uplink but not as prominent as the downlink case. FSO can be further beneficial for the downlink from the energy efficiency standpoint due to the reduced energy consumption \cite{ZediniSatelliteTWC2020, arienzo_green_2019}. However, more efforts are required to develop advanced pointing, acquisition, and tracking mechanisms (PAT) for FSO systems for near-Earth application \cite{liu_outage_2020, siddharth_outage_2020,SalmanTCOM,SalmanOJCOMS}.\newline
We note that many satellite and aerial network configurations based on hybrid RF/FSO links have been proposed in the literature, such as the following:
\begin{enumerate}
    \item Satellite-aerial-terrestrial network, where a high-altitude platform (HAP) acts as a relay between a satellite and terrestrial terminals \cite{huang_uplink_2021, swaminathan_performance_2020, kong_multiuser_2020, lee_integrating_2020}.
	\item Spatial networks, where the communications links such as satellite-to-satellite (in the same or different orbits), HAP-to-HAP. The links in these layers can be established using both RF or FSO \cite{saeed_point--point_2020}.
    \item Hybrid networks as the  RF link is for the ground-to-air channel, and the FSO link is for the air-to-air
path, and a hybrid RF/FSO system for the air-to-ground channel. Such a setup utilizes the frequency spectrum efficiently, enhances data rates, and provides inherent security \cite{erdogan_cognitive_2020}.
    \item Multibeam high-throughput satellite systems where the feeder link operates in the FSO band and the user link, between the satellite and the user terminal, is in the RF Ka-band \cite{ahmad_next-generation_2020-1}.
\end{enumerate}

\indent Another potential application of using FSO together with RF systems that worth mentioning in this section is achieving satellite-based quantum key distribution (QKD). Satellite-based QKD consists of distributing secret encryption keys to enhance the security of long-range communication systems. Earth-Satellite QKD demonstrations over distances beyond 1000 km have been reported  \cite{SatelliteQKD,SatelliteTeleportation}. The field of satellite-based QKD is continuously growing \cite{ProgressSatQKD,CubeSatreview}, and these research and development efforts will result in a global QKD network that may enhance the encryption of conventional near-Earth communication.
\subsection{RF/FSO for Deep Space Operations}
For deep space communication, RF can benefit from the well-established infrastructure of the current systems. FSO can be used as a complementary technology to push the data rates for this infrastructure, opening the door to various applications such as outer-space real-time video streaming and the transfer of high-resolution images. Indeed, recent demonstrations have shown that optical wireless communication can potentially bring fiber-like connectivity to the space \cite{Lunar,LADEE}.
In Fall 2013, Nasa's Lunar laser communication demonstration (LLCD), which was part of the LADEE (Lunar Atmosphere and Dust Environment Explorer) mission, reported a series of full-duplex communications between a satellite in lunar orbit (400.000 from the Earth) and multiple ground stations (in Spain and US) with a maximum uplink throughput of 20 Mbits and downlink throughput of 622 Mbits \cite{Lunar,LADEE}.
In collaboration with SpaceX company, NASA is also currently working to install an FSO system for the Psyche mission to explore a metal asteroid orbiting between March and Jupiter \cite{Psyche}. The idea is to improve communication performance by 10 to 100 times over the current RF technology without incurring increases in mass, volume, or power. The FSO link is expected to be operational in 2026.
\section{Open Issues and Future Research Directions}
\label{sec:Direction}
In the following, we discuss the open issues associated with the RF/FSO co-existence, related mainly to the cost of technology deployment. We then provide a set of future and insightful research directions that we believe can potentially lead to the wide-scale deployment of parallel RF/FSO links.
\subsection{Economic Challenges}
The main limiting factor of the wide-scale deployment of RF/FSO is the cost. The cost of building the optical wireless infrastructure by network operators is a significant challenge, in particular when coherent detection is used or sophisticated PAT systems are involved for the FSO links. However, the return of investment in terrestrial applications can be quick, mainly if IM/DD systems are installed, and the FSO system is the primary link to preserve the power consumption. The cost may also vary depending on the installation's geographical location. For example, suppose the hybrid system is installed in a site where fog is a dominant weather condition. In that case, it will mostly rely on radio technology, which derives higher Operating expenses (OPEX). Programming the devices to perform the switching mechanisms can also generate extra costs.
To build confidence in such a technology, it should be installed in pilot locations such as cities with clear weather conditions throughout the year.
\subsection{Boosting the Role of Machine Learning}
Using machine learning algorithms for efficient switching between FSO and RF links has not yet been fully exploited. Efficient switching ML-based algorithms can be built based on several parameters like channel impairments. Such parameters could be predicted using various ML tools with high accuracy, as reported in \cite{Ragheb,Li:18,EsmailOPEX21}. A recent study reported in \cite{EsmailOPEX21}, has shown that turbulence, pointing errors, and optical SNR can be predicted accurately at different FSO transfer rates. 
ML-based switching can also be set on power consumption or other non-channel-related parameters such as the EMF radiation. This scenario can be suitable for cases when the RF is the primary source of communication. The FSO system,  in this case, can be turned ``on'' to reduce the amount (or duration) of EMF radiations in a particular location. \newline
\indent We should note that the adoption of ML algorithms in parallel links requires channel measurements on various RF and FSO channel conditions to build the learning models. The relatively newly introduced Generative Adversarial Networks (GAN) ML framework can be used for data augmentation in such cases. We note that GAN is a deep neural network that learns to generate new data with the same statistics of the learning set \cite{GANGoodfellow}. Using GAN can minimize the amount of outdoor collected data required to train the learning models for the RF/FSO ML-based switching method; in particular, that measurement collection can be a time-consuming process. A recent study revealed that GAN-based modeling is promising in wireless RF communication \cite{GANYang}.  

\subsection{Connecting the Unconnected}
As of January 2021, more than 35\% of the world population does not have access to the internet \cite{Stat1}. A large percentage of the connected population is also under-connected with limited access to the internet, even in developed countries.
RF/FSO hybrid backhauling can significantly contribute to connecting the unconnected and under-connected world regions. Setting new optical fiber networks can be costly, mainly in terms of Capital Expenditure (CAPEX), which involves the initial installation cost \cite{YaacoubProcIEEE}. 
OPEX of RF/FSO systems is comparable to those of RF-only systems, especially if the FSO link is operational over long hours, which reduces the overall energy consumption. Using only RF links may also not be enough to fulfill the continuously increasing capacity demand. Therefore, co-installing RF and FSO technologies can be a practical solution yet economically profitable to connect far-flung areas with higher resistance to natural disasters. In particular, commercially available \textit{plug-and-play} hybrid systems can cover distances up to 10 km and can be installed quickly without incurring civil engineering costs to lay optical fibers underground \cite{TrichiliJOSAB20}. RF/FSO system can equally contribute to enhancing the internet throughput for those living in densely populated areas and suffering from low internet penetration. Providing reliable connectivity to these locations can enable new technologies for smart agriculture and also climate change monitoring and internet of things (IoT) devices \cite{YaacoubIoTMag}. 
\subsection{Massive MIMO Hybrid RF/FSO}
Massive MIMO configuration can significantly contribute to increasing the data rates of radio links \cite{MMIMO}. A massive MIMO configuration can enable to match the throughput of RF links to what could be provided by a single FSO link. Progress in VCSEL technology has also led to the development of many FSO sources that could be modulated independently on a single chip. There are also various designs of multi-aperture detectors in direct and coherent detection configurations. Newly developed antennas can be incorporated to extend the field of view of high-speed optical detectors \cite{AlkhazragiOL21}. MIMO RF/FSO can equally improve the reliability of both links.\newline
\subsection{Non-Conventional RF/FSO hybrid systems}
There has been considerable interest in using the orbital angular momentum (OAM) degree of freedom of the electromagnetic wave to push the transmission capacity \cite{WillnerAOP15}. OAM is associated with a helical phase- front of $\exp(i\ell\phi)$ \cite{AllenOAM}, with $\ell$ being the topological charge and $\phi$ is the azimuth. The orthogonality between OAM beams enabled an extra multiplexing scheme in optical communication \cite{GibsonOE04}. OAM multiplexing is possible in FSO, as well as in radio communication \cite{Tamburini2012}. 
Beyond Tbps transmissions were reported using OAM multiplexing using optical signals propagating in free space \cite{JWangTbpsFSO,100TbpsOAM} and Multi-Gbps transmissions have also been demonstrated in the RF domain \cite{MMWOAM}. OAM multiplexing is a particular case of spatial mode multiplexing, which also involves using orthogonal mode families to carry independent data streams, such as the Laguerre Gaussian (LG), Hermite Gaussian (HG), and Ince Gaussian (IG) mode bases. Another promising aspect of using spatial mode multiplexing is fulfilling diversity without any requirements on the separation between the transmitting or receiving apertures \cite{cox2018diversity}. Diversity is not restricted to OAM-only modes derived from LG mode basis but can cover modes from different mode bases, such as LG and HG modes.\newline Djordjevic proposed in \cite{OAMFSOTHz} using OAM FSO/RF  to increase the spectral efficiency and PLS of the optical and RF wireless systems \cite{OAMFSOTHz}.\newline
\indent This non-conventional RF/FSO spatial-mode based is indeed promising. However, there are still some issues mainly associated with the design of compact mode generation and detection techniques that need to be addressed before deploying such a method beyond laboratory-test benches either in the optical or RF bands.
 \newline

\subsection{Toward Multiband Transmission}
 The THz band has recently attracted considerable attention and seems to offer plenty of opportunities for wireless communication  \cite{ElayanTHz}. Progress in devices and modeling may bring THz communication to life sooner than we thought \cite{THzNatElec18 ,HadiSurveyTHz}. Incorporating this frequency band, ranging from 100 GHz to 10 THz, fills the gap between the RF and optical bands and leads to creating multiband systems. Multiband systems can provide further resilience to a wide range of channel conditions \cite{THzResilience}. For example, as we discussed in section \ref{subsec:ChannelEffects}, FSO signals are severely affected by dust storms. However, experimental investigations, reported in \cite{THzResilience}, showed that THz signals are barely affected by dust signals. Multiband can equally open up new applications. Although water vapor in the atmosphere severely absorbs THz signals, they can be used with RF/FSO systems to provide high-throughput connectivity in non-line-of-sight scenarios or short-range transmissions.\newline
\indent Multiband transmission can be further of interest for near-Earth and deep space applications. THz beams have lower divergence compared to RF ones, which can provide a good compromise between pointing sensitivity and geometrical loss. It has been pointed out that THz communication systems can offer a comparable performance to FSO systems for Earth-to-satellite links in dry regions \cite{THzSatellite}.\newline
\indent Another advantage of parallel multiband transmissions is strengthening the secrecy of the system. If an intruder is detected in one band, switching the communication on one of the other two technologies makes signal interception or jamming difficult. It is also possible to divide the same information on the various communication bands, restricting any eavesdropper from revealing any useful information by eavesdropping on a single band. This can provide extra security to RF/FSO networks.

\subsection{Toward Self-Powered RF/FSO Networks}
Solar-powered RF base stations are being deployed in various areas of the world \cite{SolarPoweredRF}. Powering RF sites using solar panels can ensure power delivery to radio base stations in remote locations or regions without reliable grid connectivity \cite{SolarPoweredRF}. A future idea to explore is to build hybrid RF/FSO self-powered networks. A particular scenario of interest is to use FSO systems to lower power consumption in battery-powered sites to avoid or reduce the power outage period, mainly because FSO systems typically consume less energy than the RF ones, and therefore ensuring longer battery lives.\\
\indent Using solar power as an energy source can reduce the carbon footprints of cellular base stations compared to fuel-powered ones. It has also been demonstrated that off-the-shelf solar panels can be used to decode information signals while harvesting energy \cite{HaasEnergies}. Indeed, high data rates in the order of 1 Gbps can be reached when using solar cells operating as photodetectors \cite{HaasGaAs}. MIMO configuration using solar cells has been shown to be also possible in \cite{MIMOSLIPT}. Using solar cells initially used to power base station batteries to decode optical information signals without using any dedicated photodetectors is another idea to explore, mainly in scenarios when no large amount of data is transmitted. Using large area solar detectors can equally ease the PAT requirements for FSO systems even in the presence of atmospheric turbulence or slight pointing errors.  \\ 

\section{Conclusion}
\label{Conclusion}
 FSO communication is paving the way toward high-speed wireless backhauls. However, the sensitivity of optical signals to atmospheric turbulence may limit the potential of large-scale deployment of such a technology. By retrofitting FSO systems to existing RF infrastructure, it is possible to benefit from the high data rates of FSO systems and the resilience of RF links to atmospheric turbulence. Throughout this paper, we identified the motivation of using RF and FSO communication links together. We presented the different operation and switching mechanisms adopted in RF/FSO parallel links. The characteristics of various commercially available RF/FSO solutions as well as reported field trials are reviewed. The potential of using FSO on top of RF systems for near-Earth and deep space applications is highlighted. Open issues associated with hybrid RF/FSO systems' hardware and deployment cost are discussed. We finally identified future research directions that could be of interest to academics and industrials working on the design of RF/FSO parallel links. Among such directions, using RF/FSO communication can contribute to the efforts willing to connect the unconnected and enhance internet throughput in under-connected regions. Building multiband systems covering the THz band can create a wide range of new opportunities and further resilience to atmospheric conditions for terrestrial and satellite communications. RF/FSO networks can lower power consumption and, therefore, decrease outage risks in battery-powered base stations in difficult-access locations.
\section*{List of Abbreviations}
\noindent 5G: Fifth Generation\newline
6G: Sixth Generation\newline
AF: Amplify and Forward\newline
APD: Avalanche Photodetector\newline
BER: Bit Error Rate\newline
CAPEX: Capital Expenditure\newline
CSI: Channel State Information\newline
DF: Decode and Forward\newline 
DSP: Digital Signal Processing\newline
EGC: Equal Gain Combining\newline
EMF: Electromagnetic Field\newline
FSO: Free Space Optics\newline
GAN: Generative Adversarial Network\newline
GEO: Geosynchronous Earth Orbit\newline
HAP: High Altitude Platform\newline
HG: Hermite Gaussian\newline 
HTS: High Throughput Satellite\newline
IoT: Internet of Things\newline
IM/DD: Intensity Modulation/Direct Detection\newline
IR: Infrared\newline
LEO: Low Earth Satellite\newline
LG: Laguerre Gaussian\newline
LiFi: Light Fidelity\newline
LoS: Line of Sight\newline
LSTM: Long Short Term Memory\newline
MEO: Medium Earth Orbit\newline
MIMO: Multiple-Input Multiple-Output\newline
ML: Machine Learning\newline
MMW: Millimeter Wave\newline
MRC: Maximum Ratio Combining\newline
NLoS: Non-line-of sight\newline
NR: New Radio\newline
OAM: Orbital Angular Momentum\newline
OOK: On Off Keying\newline
OPEX: Operating Expenses\newline
PAT: Pointing, Acquisition, and Tracking\newline
PH: Power Hysteresis\newline
PLS: Physical Layer Security\newline
PSK: Phase Shift Keying\newline
QAM: Quadrature Amplitude Modulation\newline
QE: Quantize and Encode\newline
QKD: Quantum Key Distribution\newline
QoS: Quality of Service\newline
RF: Radio Frequency\newline
Rx: Receiver\newline
RSSI: Received Signal Strength Indicator\newline
SATN: Satellite-Aerial-Terrestrial Network\newline
SC: Selective Combining\newline
SER: Symbol Error Rate\newline
SFP: Small Form-factor Pluggable\newline
SI: Scintillation Index\newline
SIM: Sub-intensity Modulation\newline
SISO: Single-Input Single Output\newline
SNR: Signal to Noise Ratio\newline
TH: Time Hysteresis \newline
Tx: Transmitter\newline
UAV: Unmanned Aerial Vehicle\newline
VCSEL: Vertical Cavity Surface Emitting Laser\newline
VLC: Visible Light Communication\newline
WDM: Wavelength Division Multiplexing\newline
WiFi: Wireless Fidelity\newline
WOC: Wireless Optical Communication
\section*{Funding}
KAUST-KSU special initiative. KAUST office of sponsored research (OSR).
\section*{Acknowledgment}
Figure \ref{fig:FSOUseCases} was produced by Heno Hwang, a scientific illustrator at King Abdullah University of Science and Technology (KAUST).
\ifCLASSOPTIONcaptionsoff
  \newpage
\fi

\bibliographystyle{IEEEtran}
\bibliography{Main.bib}

\begin{thebibliography}{100}
\providecommand{\url}[1]{#1}
\csname url@samestyle\endcsname
\providecommand{\newblock}{\relax}
\providecommand{\bibinfo}[2]{#2}
\providecommand{\BIBentrySTDinterwordspacing}{\spaceskip=0pt\relax}
\providecommand{\BIBentryALTinterwordstretchfactor}{4}
\providecommand{\BIBentryALTinterwordspacing}{\spaceskip=\fontdimen2\font plus
\BIBentryALTinterwordstretchfactor\fontdimen3\font minus
  \fontdimen4\font\relax}
\providecommand{\BIBforeignlanguage}[2]{{%
\expandafter\ifx\csname l@#1\endcsname\relax
\typeout{** WARNING: IEEEtran.bst: No hyphenation pattern has been}%
\typeout{** loaded for the language `#1'. Using the pattern for}%
\typeout{** the default language instead.}%
\else
\language=\csname l@#1\endcsname
\fi
#2}}
\providecommand{\BIBdecl}{\relax}
\BIBdecl

\bibitem{FSOBackhauling}
M.~{Alzenad}, M.~Z. {Shakir}, H.~{Yanikomeroglu}, and M.-S. {Alouini},
  ``{FSO}-based vertical backhaul/fronthaul framework for {5G+} wireless
  networks,'' \emph{IEEE Communications Magazine}, vol.~56, no.~1, pp.
  218--224, 2018.

\bibitem{ChenCommag20}
Y.~W. {Chen}, R.~{Zhang}, C.~W. {Hsu}, and G.~K. {Chang}, ``Key enabling
  technologies for the post-{5G} era: {Fully} adaptive, all-spectra coordinated
  radio access network with function decoupling,'' \emph{IEEE Communications
  Magazine}, vol.~58, no.~9, pp. 60--66, 2020.

\bibitem{6GVTMag}
K.~{David} and H.~{Berndt}, ``{6G} vision and requirements: {Is} there any need
  for beyond {5G?}'' \emph{IEEE Vehicular Technology Magazine}, vol.~13, no.~3,
  pp. 72--80, 2018.

\bibitem{6GNatureElectro}
S.~{Dang}, O.~{Amin}, B.~{Shihada}, and M.-S. {Alouini}, ``What should {6G}
  be?'' \emph{Nature Electronics}, vol.~3, p. 20–29, 2020.

\bibitem{TrichiliConst19}
A.~{Trichili}, K.~{Park}, M.~{Zghal}, B.~S. {Ooi}, and M.-S. {Alouini},
  ``Communicating using spatial mode multiplexing: Potentials, challenges, and
  perspectives,'' \emph{IEEE Communications Surveys \& Tutorials}, vol.~21,
  no.~4, pp. 3175--3203, 2019.

\bibitem{FSOWDM}
G.~Nykolak, P.~F. Szajowski, J.~Jacques, H.~M. Presby, J.~A. Abate, G.~E.
  Tourgee, and J.~Auborn, ``$4\times2.5$ {G}b/s 4.4 km {WDM} free-space optical
  link at 1550 nm,'' in \emph{Optical Fiber Communication Conference and the
  International Conference on Integrated Optics and Optical Fiber
  Communication}.\hskip 1em plus 0.5em minus 0.4em\relax Optical Society of
  America, 1999, p. PD11.

\bibitem{CvijeticPDM10}
N.~Cvijetic, D.~Qian, J.~Yu, Y.-K. Huang, and T.~Wang,
  ``Polarization-multiplexed optical wireless transmission with coherent
  detection,'' \emph{J. Lightwave Technol.}, vol.~28, no.~8, pp. 1218--1227,
  Apr 2010.

\bibitem{KimSpie11}
I.~I. Kim and E.~J. Korevaar, ``{Availability of free-space optics (FSO) and
  hybrid FSO/RF systems},'' in \emph{Optical Wireless Communications IV}, E.~J.
  Korevaar, Ed., vol. 4530, International Society for Optics and
  Photonics.\hskip 1em plus 0.5em minus 0.4em\relax SPIE, 2001, pp. 84 -- 95.

\bibitem{FiberWithoutFiber}
H.~A. {Willebrand} and B.~S. {Ghuman}, ``Fiber optics without fiber,''
  \emph{IEEE Spectrum}, vol.~38, no.~8, pp. 40--45, 2001.

\bibitem{DouikIEEETCOM16}
A.~{Douik}, H.~{Dahrouj}, T.~Y. {Al-Naffouri}, and M.~{Alouini}, ``Hybrid
  radio/free-space optical design for next generation backhaul systems,''
  \emph{IEEE Transactions on Communications}, vol.~64, no.~6, pp. 2563--2577,
  2016.

\bibitem{UysalComst14}
M.~A. Khalighi and M.~Uysal, ``Survey on free space optical communication: A
  communication theory perspective,'' \emph{IEEE communications surveys \&
  tutorials}, vol.~16, no.~4, pp. 2231--2258, 2014.

\bibitem{GhassemlooyJSAC15}
Z.~{Ghassemlooy}, S.~{Arnon}, M.~{Uysal}, Z.~{Xu}, and J.~{Cheng}, ``Emerging
  optical wireless communications-advances and challenges,'' \emph{IEEE Journal
  on Selected Areas in Communications}, vol.~33, no.~9, pp. 1738--1749, 2015.

\bibitem{Space_FSO}
H.~Kaushal and G.~Kaddoum, ``Optical communication in space: Challenges and
  mitigation techniques,'' \emph{IEEE Communications Surveys \& Tutorials},
  vol.~19, no.~1, pp. 57--96, 2017.

\bibitem{TrichiliJOSAB20}
A.~Trichili, M.~A. Cox, B.~S. Ooi, and M.-S. Alouini, ``Roadmap to free space
  optics,'' \emph{Journal of the Optical Society of America B}, vol.~37,
  no.~11, pp. A184--A201, Nov 2020.

\bibitem{OWHybridNetworks}
M.~Z. {Chowdhury}, M.~K. {Hasan}, M.~{Shahjalal}, M.~T. {Hossan}, and Y.~M.
  {Jang}, ``Optical wireless hybrid networks: Trends, opportunities,
  challenges, and research directions,'' \emph{IEEE Communications Surveys \&
  tutorials}, vol.~22, no.~2, pp. 930--966, 2020.

\bibitem{WiFiLiFi}
X.~Wu, M.~D. Soltani, L.~Zhou, M.~Safari, and H.~Haas, ``Hybrid {LiFi} and
  {WiFi} networks: A survey,'' \emph{IEEE Communications Surveys \& Tutorials},
  2021.

\bibitem{RFHealthRisks}
L.~Chiaraviglio, A.~Elzanaty, and M.-S. Alouini, ``Health risks associated with
  {5G} exposure: A view from the communications engineering perspective,''
  \emph{arxiv preprint}, 2020, arXiv:2006.00944.

\bibitem{fsonaSecrecy}
{FSONA Optical Wireless}, ``Application note - {Security} of a free space
  optical transmission,'' http://www.fsona.com/tech/app\textunderscore
  notes/fSONA-APPNOTE-FSO\textunderscore Security.pdf (Acessed 8 February
  2021).

\bibitem{SecrecyRF-FSO}
Y.~Ai, A.~Mathur, H.~Lei, M.~Cheffena, and I.~S. Ansari, ``Secrecy enhancement
  of {RF} backhaul system with parallel {FSO} communication link,''
  \emph{Optics Communications}, vol. 475, p. 126193, 2020.

\bibitem{LopezMartinexPJ15}
F.~J. {Lopez-Martinez}, G.~{Gomez}, and J.~M. {Garrido-Balsells},
  ``Physical-layer security in free-space optical communications,'' \emph{IEEE
  Photonics Journal}, vol.~7, no.~2, pp. 1--14, 2015.

\bibitem{Rayleighscattering}
A.~T. Young, ``Rayleigh scattering,'' \emph{Applied Optics}, vol.~20, no.~4,
  pp. 533--535, Feb 1981.

\bibitem{7515199}
M.~A. {Esmail}, H.~{Fathallah}, and M.-S. {Alouini}, ``Outdoor {FSO}
  communications under fog: Attenuation modeling and performance evaluation,''
  \emph{IEEE Photonics Journal}, vol.~8, no.~4, pp. 1--22, 2016.

\bibitem{7501899}
------, ``An experimental study of {FSO} link performance in desert
  environment,'' \emph{IEEE Communications Letters}, vol.~20, no.~9, pp.
  1888--1891, 2016.

\bibitem{AndrewsBook}
L.~C. Andrews and R.~L. Phillips, \emph{Laser Beam Propagation through Random
  Media}, 2nd~ed.\hskip 1em plus 0.5em minus 0.4em\relax SPIE Press, 2005.

\bibitem{PointingErrors}
D.~{Kedar} and S.~{Arnon}, ``Urban optical wireless communication networks: the
  main challenges and possible solutions,'' \emph{IEEE Communications
  Magazine}, vol.~42, no.~5, pp. S2--S7, 2004.

\bibitem{PointingModel}
A.~A. {Farid} and S.~{Hranilovic}, ``Outage capacity optimization for
  free-space optical links with pointing errors,'' \emph{Journal of Lightwave
  Technology}, vol.~25, no.~7, pp. 1702--1710, 2007.

\bibitem{kim}
I.~I. Kim, B.~McArthur, and E.~J. Korevaar, ``{Comparison of laser beam
  propagation at 785 nm and 1550 nm in fog and haze for optical wireless
  communications},'' in \emph{Optical Wireless Communications III}, E.~J.
  Korevaar, Ed., vol. 4214, International Society for Optics and
  Photonics.\hskip 1em plus 0.5em minus 0.4em\relax SPIE, 2001, pp. 26 -- 37.

\bibitem{mageddust}
M.~A. {Esmail}, H.~{Fathallah}, and M.-S. {Alouini}, ``An experimental study of
  {FSO} link performance in desert environment,'' \emph{IEEE Communications
  Letters}, vol.~20, no.~9, pp. 1888--1891, 2016.

\bibitem{snow}
M.~S. Awan, L.~Csurgai-Horv{\'a}th, S.~S. Muhammad, E.~Leitgeb, F.~Nadeem, and
  M.~S. Khan, ``Characterization of fog and snow attenuations for free-space
  optical propagation,'' \emph{Journal of Communications}, vol.~4, no.~8, pp.
  533--545, 2009.

\bibitem{rain}
A.~Suriza, I.~{Md Rafiqul}, A.~Wajdi, and A.~Naji, ``Proposed parameters of
  specific rain attenuation prediction for free space optics link operating in
  tropical region,'' \emph{Journal of Atmospheric and Solar-Terrestrial
  Physics}, vol.~94, pp. 93 -- 99, 2013.

\bibitem{lognormal}
Z.~N. {Chaleshtory}, A.~{Gholami}, Z.~{Ghassemlooy}, and M.~{Sedghi},
  ``Experimental investigation of environment effects on the {FSO} link with
  turbulence,'' \emph{IEEE Photonics Technology Letters}, vol.~29, no.~17, pp.
  1435--1438, 2017.

\bibitem{MalagaDistribution}
A.~Jurado-Navas, J.~M. Garrido-Balsells, J.~F. Paris, and A.~Puerta-Notario,
  \emph{Numerical Simulations of Physical and Engineering Processes}.\hskip 1em
  plus 0.5em minus 0.4em\relax IntechOpen, 2011, ch. A Unifying Statistical
  Model for Atmospheric Optical Scintillation.

\bibitem{slimbook}
M.~K. Simon and M.-S. Alouini, \emph{Digital communication over fading
  channels}.\hskip 1em plus 0.5em minus 0.4em\relax John Wiley \& Sons, 2005,
  vol.~95.

\bibitem{RiceFading}
A.~{Abdi}, C.~{Tepedelenlioglu}, M.~{Kaveh}, and G.~{Giannakis}, ``On the
  estimation of the {K} parameter for the {Rice} fading distribution,''
  \emph{IEEE Communications Letters}, vol.~5, no.~3, pp. 92--94, 2001.

\bibitem{RFFSOLee}
E.~{Lee}, J.~{Park}, D.~{Han}, and G.~{Yoon}, ``Performance analysis of the
  asymmetric dual-hop relay transmission with mixed {RF/FSO} links,''
  \emph{IEEE Photonics Technology Letters}, vol.~23, no.~21, pp. 1642--1644,
  2011.

\bibitem{RFFSOSamimi}
H.~{Samimi} and M.~{Uysal}, ``End-to-end performance of mixed {RF/FSO}
  transmission systems,'' \emph{IEEE/OSA Journal of Optical Communications and
  Networking}, vol.~5, no.~11, pp. 1139--1144, 2013.

\bibitem{ZediniPJ15}
E.~{Zedini}, I.~S. {Ansari}, and M.~{Alouini}, ``Performance analysis of mixed
  {Nakagami}-$m$ and {Gamma–Gamma} dual-hop {FSO} transmission systems,''
  \emph{IEEE Photonics Journal}, vol.~7, no.~1, pp. 1--20, 2015.

\bibitem{AneesIET15}
S.~Anees and M.~R. Bhatnagar, ``Performance evaluation of decode-and-forward
  dual-hop asymmetric radio frequency-free space optical communication
  system,'' \emph{IET Optoelectronics}, vol.~9, no.~5, p. 232 – 240, Oct
  2015.

\bibitem{DualHopGeneralizedRFFSO}
E.~{Zedini}, H.~{Soury}, and M.-S. {Alouini}, ``On the performance analysis of
  dual-hop mixed {FSO/RF} systems,'' \emph{IEEE Transactions on Wireless
  Communications}, vol.~15, no.~5, pp. 3679--3689, 2016.

\bibitem{SIMModulation}
W.~O. {Popoola} and Z.~{Ghassemlooy}, ``{BPSK} subcarrier intensity modulated
  free-space optical communications in atmospheric turbulence,'' \emph{Journal
  of Lightwave Technology}, vol.~27, no.~8, pp. 967--973, 2009.

\bibitem{OutdatedCSI}
G.~T. {Djordjevic}, M.~I. {Petkovic}, A.~M. {Cvetkovic}, and G.~K.
  {Karagiannidis}, ``Mixed {RF/FSO} relaying with outdated channel state
  information,'' \emph{IEEE Journal on Selected Areas in Communications},
  vol.~33, no.~9, pp. 1935--1948, 2015.

\bibitem{QuantizeEncode}
K.~{Kumar} and D.~K. {Borah}, ``Quantize and encode relaying through {FSO} and
  hybrid {FSO/RF} links,'' \emph{IEEE Transactions on Vehicular Technology},
  vol.~64, no.~6, pp. 2361--2374, 2015.

\bibitem{UsmanPJ14}
M.~{Usman}, H.~{Yang}, and M.-S. {Alouini}, ``Practical switching-based hybrid
  {FSO/RF} transmission and its performance analysis,'' \emph{IEEE Photonics
  Journal}, vol.~6, no.~5, pp. 1--13, 2014.

\bibitem{VISHWAKARMA2021126796}
N.~Vishwakarma and S.~R, ``Performance analysis of hybrid {FSO/RF}
  communication over generalized fading models,'' \emph{Optics Communications},
  vol. 487, p. 126796, 2021.

\bibitem{AbadiICC17}
M.~M. {Abadi}, Z.~{Ghassemlooy}, S.~{Zvanovec}, M.~R. {Bhatnagar}, and Y.~{Wu},
  ``Hard switching in hybrid {FSO/RF} link: Investigating data rate and link
  availability,'' in \emph{2017 IEEE International Conference on Communications
  Workshops (ICC Workshops)}, 2017, pp. 463--468.

\bibitem{RakiaTWC20}
T.~{Rakia}, F.~{Gebali}, H.~{Yang}, and M.-S. {Alouini}, ``Performance analysis
  of multiuser {FSO/RF} network under non-equal priority with $p$ -persistence
  protocol,'' \emph{IEEE Transactions on Wireless Communications}, vol.~19,
  no.~3, pp. 1802--1813, 2020.

\bibitem{Sharma:19}
S.~{Sharma}, A.~S. {Madhukumar}, and S.~{R}, ``Switching-based cooperative
  decode-and-forward relaying for hybrid {FSO/RF} networks,'' \emph{Journal of
  Optical Communications and Networking}, vol.~11, no.~6, pp. 267--281, Jun
  2019.

\bibitem{ZhangJSAC09}
W.~{Zhang}, S.~{Hranilovic}, and C.~{Shi}, ``Soft-switching hybrid {FSO/RF}
  links using short-length raptor codes: Design and implementation,''
  \emph{IEEE Journal on Selected Areas in Communications}, vol.~27, no.~9, pp.
  1698--1708, 2009.

\bibitem{TangTCOM12}
Y.~{Tang}, M.~{Brandt-Pearce}, and S.~G. {Wilson}, ``Link adaptation for
  throughput optimization of parallel channels with application to hybrid
  {FSO/RF} systems,'' \emph{IEEE Transactions on Communications}, vol.~60,
  no.~9, pp. 2723--2732, 2012.

\bibitem{9184046}
W.~S. {Saif}, M.~A. {Esmail}, A.~M. {Ragheb}, T.~A. {Alshawi}, and S.~A.
  {Alshebeili}, ``Machine learning techniques for optical performance
  monitoring and modulation format identification: A survey,'' \emph{IEEE
  Communications Surveys \& Tutorials}, vol.~22, no.~4, pp. 2839--2882, 2020.

\bibitem{8527529}
F.~{Musumeci}, C.~{Rottondi}, A.~{Nag}, I.~{Macaluso}, D.~{Zibar},
  M.~{Ruffini}, and M.~{Tornatore}, ``An overview on application of machine
  learning techniques in optical networks,'' \emph{IEEE Communications Surveys
  \& Tutorials}, vol.~21, no.~2, pp. 1383--1408, 2019.

\bibitem{8666641}
C.~{Zhang}, P.~{Patras}, and H.~{Haddadi}, ``Deep learning in mobile and
  wireless networking: A survey,'' \emph{IEEE Communications Surveys \&
  Tutorials}, vol.~21, no.~3, pp. 2224--2287, 2019.

\bibitem{haluvska2020}
R.~Halu{\v{s}}ka, P.~{\v{S}}ul'aj, L.~Ovsen{\'\i}k, S.~Marchevsk{\`y},
  J.~Papaj, and L.~Dobo{\v{s}}, ``Prediction of received optical power for
  switching hybrid {FSO/RF} system,'' \emph{Electronics}, vol.~9, no.~8, p.
  1261, 2020.

\bibitem{toth2018c}
J.~T{\'o}th, L.~Ovsen{\'\i}k, J.~Tur{\'a}n, L.~Michaeli, and M.~M{\'a}rton,
  ``Classification prediction analysis of {RSSI} parameter in hard switching
  process for {FSO/RF} systems,'' \emph{Measurement}, vol. 116, pp. 602--610,
  2018.

\bibitem{meng2019}
Y.~Meng, Y.~Liu, S.~Song, Y.~Yang, and L.~Guo, ``Predictive link switching for
  energy efficient {FSO/RF} communication system,'' in \emph{Proc. of Asia
  Communications and Photonics Conference (ACP)}, 2019, pp. 1--3.

\bibitem{FSORFDiversity}
N.~D. {Chatzidiamantis}, G.~K. {Karagiannidis}, E.~E. {Kriezis}, and
  M.~{Matthaiou}, ``Diversity combining in hybrid {RF/FSO} systems with {PSK}
  modulation,'' in \emph{2011 IEEE International Conference on Communications
  (ICC)}, 2011, pp. 1--6.

\bibitem{AMIRABADI2019293}
M.~A. Amirabadi and V.~{Tabataba Vakili}, ``A novel hybrid {FSO / RF}
  communication system with receive diversity,'' \emph{Optik}, vol. 184, pp.
  293 -- 298, 2019.

\bibitem{RakiaPJ15}
T.~{Rakia}, H.~{Yang}, F.~{Gebali}, and M.-S. {Alouini}, ``Power adaptation
  based on truncated channel inversion for hybrid {FSO/RF} transmission with
  adaptive combining,'' \emph{IEEE Photonics Journal}, vol.~7, no.~4, pp.
  1--12, 2015.

\bibitem{8883295}
M.~Z. {Hassan}, M.~J. {Hossain}, J.~{Cheng}, and V.~C.~M. {Leung}, ``Hybrid
  {RF/FSO} backhaul networks with statistical-{QoS}-aware buffer-aided
  relaying,'' \emph{IEEE Transactions on Wireless Communications}, vol.~19,
  no.~3, pp. 1464--1483, 2020.

\bibitem{cbl_datasheet}
``Communication by light,'' \url{https://www.cbl.de/en/downloads/data-sheets/},
  last accessed: 22-12-2020.

\bibitem{jv-labs}
jv~labs, \url{http://www.jv-shop.eu}, last accessed: 22-12-2020.

\bibitem{teldata}
Teldata,
  \url{http://www.teldata.net/teldata-datasheets/DS\_FlightStrata\_100\_XA.pdf},
  last accessed: 22-12-2020.

\bibitem{lightpointe}
LightPointe, \url{http://site.microcom.us/shl212505hs0er}, last accessed:
  22-12-2020.

\bibitem{MostCom}
``{FSO} technology,'' \url{http://www.moctkom.ru/fso-technology/}, last
  accessed 02.12.2020.

\bibitem{collinear}
Collinear, \url{https://www.collinear.com/hybridcnx/}, last accessed:
  22-12-2020.

\bibitem{ECsystems}
E.~Systems,
  \url{http://www.ecsystem.cz/en/products/free-space-optic-equipment}, last
  accessed: 22-12-2020.

\bibitem{Trimble}
``Free space optics wireless communication solutions,''
  \url{http://trimble.hu/fso\_en}, last accessed: 22-12-2020.

\bibitem{sona}
fSONA, \url{http://www.fsona.com}, last accessed: 22-12-2020.

\bibitem{hyper}
{Hyperion Technologies}, online, Accessed, Dec 2020,
  http://www.hyperiontechs.com/5g-small-cell-backhaul/.

\bibitem{anova}
ANOVA, \url{https://anovanetworks.com/celeras/}, last accessed: 22-12-2020.

\bibitem{bussinesswire}
``Global free space optics communications analysis,''
  \url{https://www.businesswire.com/news/home/20200924005531/en/Global-Free-Space-Optics-Communications-Analysis-2020-Worldwide-Commercial-Use-of-Commercial-Terrestrial-FSO-TransmitterReceivers-is-Forecast-to-Reach-363-million-in-2029---ResearchAndMarkets.com/},
  last accessed: 23-12-2020.

\bibitem{1505831}
A.~{Akbulut}, H.~G. {Ilk}, and F.~{Ari}, ``Design, availability and reliability
  analysis on an experimental outdoor {FSO/RF} communication system,'' in
  \emph{Proceedings of 2005 7th International Conference Transparent Optical
  Networks, 2005.}, vol.~1, 2005, pp. 403--406.

\bibitem{10.1117/12.2546441}
R.~B. Youssef, J.~L. Riggins, M.~P. O'Toole, K.~T. Newell, J.~W. Zobel,
  K.~Patel, J.~E. Malowicki, R.~J. Dimeo, V.~Bedi, and W.~G. Cook, ``{Hybrid
  FSO/RF communications system for high-availability, high-capacity
  networks},'' in \emph{Free-Space Laser Communications XXXII}, H.~Hemmati and
  D.~M. Boroson, Eds., vol. 11272, International Society for Optics and
  Photonics.\hskip 1em plus 0.5em minus 0.4em\relax SPIE, 2020, pp. 33 -- 40.

\bibitem{RFFSODemonstration}
B.~Rödiger, D.~Ginthör, J.~P. Labrador, J.~Ramirez, C.~Schmidt, and C.~Fuchs,
  ``{Demonstration of an FSO/RF hybrid-communication system on aeronautical and
  space applications},'' in \emph{Laser Communication and Propagation through
  the Atmosphere and Oceans IX}, J.~A. Anguita, J.~P. Bos, and D.~T. Wayne,
  Eds., vol. 11506, International Society for Optics and Photonics.\hskip 1em
  plus 0.5em minus 0.4em\relax SPIE, 2020, pp. 1 -- 11.

\bibitem{7181911}
S.~J. {Hussain}, A.~{Touati}, M.~{Elamri}, F.~{Touati}, and M.~{Uysal},
  ``Evaluation of {FSO} link throughput in {Qatar},'' in \emph{2015 4th
  Mediterranean Conference on Embedded Computing (MECO)}, 2015, pp. 232--235.

\bibitem{10.1117/12.2256792}
A.~Touati, F.~Touati, A.~Abdaoui, A.~Khandakar, S.~J. Hussain, and
  A.~Bouallegue, ``{An experimental performance evaluation of the hybrid
  FSO/RF},'' in \emph{Free-Space Laser Communication and Atmospheric
  Propagation XXIX}, H.~Hemmati and D.~M. Boroson, Eds., vol. 10096,
  International Society for Optics and Photonics.\hskip 1em plus 0.5em minus
  0.4em\relax SPIE, 2017, pp. 409 -- 415.

\bibitem{SaeedCubesat}
N.~{Saeed}, A.~{Elzanaty}, H.~{Almorad}, H.~{Dahrouj}, T.~Y. {Al-Naffouri}, and
  M.-S. {Alouini}, ``Cubesat communications: Recent advances and future
  challenges,'' \emph{IEEE Communications Surveys \& Tutorials}, vol.~22,
  no.~3, pp. 1839--1862, 2020.

\bibitem{DLProjects}
R.~M. Calvo, J.~Poliak, J.~Surof, A.~Reeves, M.~Richerzhagen, H.~F. Kelemu,
  R.~Barrios, C.~Carrizo, R.~Wolf, F.~Rein, A.~Dochhan, K.~Saucke, and
  W.~Luetke, ``{Optical technologies for very high throughput satellite
  communications},'' in \emph{Free-Space Laser Communications XXXI}, H.~Hemmati
  and D.~M. Boroson, Eds., vol. 10910, International Society for Optics and
  Photonics.\hskip 1em plus 0.5em minus 0.4em\relax SPIE, 2019, pp. 189 -- 204.

\bibitem{Hemmati}
H.~{Hemmati}, A.~{Biswas}, and I.~B. {Djordjevic}, ``Deep-space optical
  communications: Future perspectives and applications,'' \emph{Proceedings of
  the IEEE}, vol.~99, no.~11, pp. 2020--2039, 2011.

\bibitem{NFIRETSX}
R.~Fields, C.~Lunde, R.~Wong, J.~Wicker, D.~Kozlowski, J.~Jordan, B.~Hansen,
  G.~Muehlnikel, W.~Scheel, U.~Sterr, R.~Kahle, and R.~Meyer,
  ``{NFIRE-to-TerraSAR-X laser communication results: satellite pointing,
  disturbances, and other attributes consistent with successful performance},''
  in \emph{Sensors and Systems for Space Applications III}, J.~L. Cox and
  P.~Motaghedi, Eds., vol. 7330, International Society for Optics and
  Photonics.\hskip 1em plus 0.5em minus 0.4em\relax SPIE, 2009, pp. 211 -- 225.

\bibitem{NFIRETSXEDRS}
S.~{Seel}, H.~{Kämpfner}, F.~{Heine}, D.~{Dallmann}, G.~{Mühlnikel},
  M.~{Gregory}, M.~{Reinhardt}, K.~{Saucke}, J.~{Muckherjee}, U.~{Sterr},
  B.~{Wandernoth}, R.~{Meyer}, and R.~{Czichy}, ``Space to ground bidirectional
  optical communication link at 5.6 {Gbps} and {EDRS} connectivity outlook,''
  in \emph{2011 Aerospace Conference}, 2011, pp. 1--7.

\bibitem{ETSVI}
Y.~Arimoto, M.~Toyoshima, M.~Toyoda, T.~Takahashi, M.~Shikatani, and K.~Araki,
  ``{Preliminary result on laser communication experiment using Engineering
  Test Satellite-VI (ETS-VI)},'' in \emph{Free-Space Laser Communication
  Technologies VII}, G.~S. Mecherle, Ed., vol. 2381, International Society for
  Optics and Photonics.\hskip 1em plus 0.5em minus 0.4em\relax SPIE, 1995, pp.
  151 -- 158.

\bibitem{Sota}
C.~Fuchs, D.~Kolev, F.~Moll, A.~Shrestha, M.~Brechtelsbauer, F.~Rein,
  C.~Schmidt, M.~Akioka, Y.~Munemasa, H.~Takenaka, and M.~Toyoshima, ``{Sota
  optical downlinks to DLR’s optical ground stations},'' in
  \emph{International Conference on Space Optics — ICSO 2016}, B.~Cugny,
  N.~Karafolas, and Z.~Sodnik, Eds., vol. 10562, International Society for
  Optics and Photonics.\hskip 1em plus 0.5em minus 0.4em\relax SPIE, 2017, pp.
  1228 -- 1236.

\bibitem{TBIRD}
B.~S. Robinson, D.~M. Boroson, C.~M. Schieler, F.~I. Khatri, O.~Guldner,
  S.~Constantine, T.~Shih, J.~W. Burnside, B.~C. Bilyeu, F.~Hakimi, A.~Garg,
  G.~Allen, E.~Clements, and D.~M. Cornwell, ``{TeraByte InfraRed Delivery
  (TBIRD): a demonstration of large-volume direct-to-Earth data transfer from
  low-Earth orbit },'' in \emph{Free-Space Laser Communication and Atmospheric
  Propagation XXX}, H.~Hemmati and D.~M. Boroson, Eds., vol. 10524,
  International Society for Optics and Photonics.\hskip 1em plus 0.5em minus
  0.4em\relax SPIE, 2018, pp. 253 -- 258.

\bibitem{Starlink}
M.~Handley, ``Delay is not an option: Low latency routing in space,'' in
  \emph{Proceedings of the 17th ACM Workshop on Hot Topics in Networks}, ser.
  HotNets ’18.\hskip 1em plus 0.5em minus 0.4em\relax New York, NY, USA:
  Association for Computing Machinery, 2018, p. 85–91.

\bibitem{Telsat}
I.~del Portillo, B.~G. Cameron, and E.~F. Crawley, ``A technical comparison of
  three low earth orbit satellite constellation systems to provide global
  broadband,'' \emph{Acta Astronautica}, vol. 159, pp. 123 -- 135, 2019.

\bibitem{ZediniSatelliteTWC2020}
E.~{Zedini}, A.~{Kammoun}, and M.-S. {Alouini}, ``Performance of multibeam very
  high throughput satellite systems based on {FSO} feeder links with {HPA}
  nonlinearity,'' \emph{IEEE Transactions on Wireless Communications}, vol.~19,
  no.~9, pp. 5908--5923, 2020.

\bibitem{arienzo_green_2019}
L.~Arienzo, ``Green {RF}/{FSO} communications in cognitive relay-based space
  information networks for maritime surveillance,'' \emph{IEEE Transactions on
  Cognitive Communications and Networking}, vol.~5, no.~4, pp. 1182--1193, Dec.
  2019.

\bibitem{liu_outage_2020}
X.~Liu, M.~Lin, W.-P. Zhu, J.-Y. Wang, and P.~K. Upadhyay, ``Outage performance
  for mixed {FSO-RF} transmission in satellite-aerial- terrestrial networks,''
  \emph{IEEE Photonics Technology Letters}, vol.~32, no.~21, pp. 1349--1352,
  Nov 2020.

\bibitem{siddharth_outage_2020}
M.~Siddharth, S.~Shah, and S.~R, ``Outage analysis of adaptive combining scheme
  for hybrid {FSO}/{RF} communication,'' in \emph{2020 {National} {Conference}
  on {Communications} ({NCC})}, Feb. 2020, pp. 1--6.

\bibitem{SalmanTCOM}
M.~S. Bashir and M.-S. Alouini, ``Adaptive acquisition schemes for
  photon-limited free-space optical communications,'' \emph{IEEE Transactions
  on Communications}, vol.~69, no.~1, pp. 416--428, 2021.

\bibitem{SalmanOJCOMS}
M.~S. Bashir, M.-C. Tsai, and M.-S. Alouini, ``Cramér-rao bounds for beam
  tracking with photon counting detector arrays in free-space optical
  communications,'' \emph{IEEE Open Journal of the Communications Society}, pp.
  1--1, 2021.

\bibitem{huang_uplink_2021}
Q.~Huang, M.~Lin, W.-P. Zhu, J.~Cheng, and M.-S. Alouini, ``Uplink massive
  access in mixed {RF}/{FSO} satellite-aerial-terrestrial networks,''
  \emph{IEEE Transactions on Communications}, pp. 1--1, 2021.

\bibitem{swaminathan_performance_2020}
R.~Swaminathan, S.~Sharma, and A.~S. MadhuKumar, ``Performance analysis of
  {HAPS}-based relaying for hybrid {FSO}/{RF} downlink satellite
  communication,'' in \emph{2020 {IEEE} 91st {Vehicular} {Technology}
  {Conference} ({VTC2020}-{Spring})}, May 2020, pp. 1--5.

\bibitem{kong_multiuser_2020}
H.~Kong, M.~Lin, W.-P. Zhu, H.~Amindavar, and M.-S. Alouini, ``Multiuser
  scheduling for asymmetric {FSO}/{RF} links in satellite-{UAV}-terrestrial
  networks,'' \emph{IEEE Wireless Communications Letters}, vol.~9, no.~8, pp.
  1235--1239, Aug. 2020, conference Name: IEEE Wireless Communications Letters.

\bibitem{lee_integrating_2020}
J.-H. Lee, J.~Park, M.~Bennis, and Y.-C. Ko, ``Integrating {LEO} satellites and
  multi-{UAV} reinforcement learning for hybrid {FSO}/{RF} non-terrestrial
  networks,'' \emph{arXiv preprint}, Oct. 2020, arXiv: 2010.10138.

\bibitem{saeed_point--point_2020}
N.~Saeed, H.~Almorad, H.~Dahrouj, T.~Y. Al-Naffouri, J.~S. Shamma, and M.-S.
  Alouini, ``Point-to-point communication in integrated satellite-aerial
  networks: State-of-the-art and future challenges,'' \emph{arXiv preprint},
  Dec. 2020.

\bibitem{erdogan_cognitive_2020}
E.~{Erdogan}, I.~{Altunbas}, N.~{Kabaoglu}, and H.~{Yanikomeroglu}, ``A
  cognitive radio enabled {RF/FSO} communication model for aerial relay
  networks: Possible configurations and opportunities,'' \emph{IEEE Open
  Journal of Vehicular Technology}, vol.~2, pp. 45--53, 2021.

\bibitem{ahmad_next-generation_2020-1}
I.~Ahmad, K.~D. Nguyen, N.~Letzepis, and G.~Lechner, ``On the next-generation
  high throughput satellite systems with optical feeder links,'' \emph{IEEE
  Systems Journal}, pp. 1--12, 2020.

\bibitem{SatelliteQKD}
S.-K. Liao, W.-Q. Cai, W.-Y. Liu, and et~al., ``Satellite-to-ground quantum key
  distribution,'' \emph{Nature}, vol. 549, no. 43–47, 2017.

\bibitem{SatelliteTeleportation}
J.~Y. et~al., ``Satellite-based entanglement distribution over 1200
  kilometers,'' \emph{Science}, vol. 356, no. 6343, pp. 1140--1144, 2017.

\bibitem{ProgressSatQKD}
R.~Bedington, J.~M. Arrazola, and A.~Ling, ``Progress in satellite quantum key
  distribution,'' \emph{npj Quantum Information}, vol.~3, no.~30, 2017.

\bibitem{CubeSatreview}
D.~K.~L. Oi, A.~Ling, J.~A. Grieve, T.~Jennewein, A.~N. Dinkelaker, and
  M.~Krutzik, ``Nanosatellites for quantum science and technology,''
  \emph{Contemporary Physics}, vol.~58, no.~1, pp. 25--52, 2017.

\bibitem{Lunar}
D.~M. Boroson, B.~S. Robinson, D.~V. Murphy, D.~A. Burianek, F.~Khatri, J.~M.
  Kovalik, Z.~Sodnik, and D.~M. Cornwell, ``{Overview and results of the Lunar
  Laser Communication Demonstration},'' in \emph{Free-Space Laser Communication
  and Atmospheric Propagation XXVI}, H.~Hemmati and D.~M. Boroson, Eds., vol.
  8971, International Society for Optics and Photonics.\hskip 1em plus 0.5em
  minus 0.4em\relax SPIE, 2014, pp. 213 -- 223.

\bibitem{LADEE}
R.~C. Elphic, G.~T. Delory, B.~P. Hine, P.~R. Mahaffy, M.~Horanyi,
  A.~Colaprete, M.~Benna, S.~K. Noble, and {The LADEE Science Team}, ``The
  lunar atmosphere and dust environment explorer mission,'' \emph{Space Science
  Reviews}, vol. 185, 2014.

\bibitem{Psyche}
B.~R. Buck, G.~D. Allen, E.~K. Duerr, K.~A. McIntosh, S.~T. Moynihan, V.~N.
  Shukla, and J.~D. Wang, ``{Photon counting camera for the NASA deep space
  optical communication demonstration on the PSYCHE mission},'' in
  \emph{Advanced Photon Counting Techniques XIII}, M.~A. Itzler, J.~C.
  Bienfang, and K.~A. McIntosh, Eds., vol. 10978, International Society for
  Optics and Photonics.\hskip 1em plus 0.5em minus 0.4em\relax SPIE, 2019, pp.
  30 -- 40.

\bibitem{Ragheb}
A.~Ragheb, W.~Saif, A.~Trichili, I.~Ashry, M.~A. Esmail, M.~Altamimi,
  A.~Almaiman, E.~Altubaishi, B.~S. Ooi, M.-S. Alouini, and S.~Alshebeili,
  ``Identifying structured light modes in a desert environment using machine
  learning algorithms,'' \emph{Optics Express}, vol.~28, no.~7, pp. 9753--9763,
  Mar 2020.

\bibitem{Li:18}
J.~Li, M.~Zhang, D.~Wang, S.~Wu, and Y.~Zhan, ``Joint atmospheric turbulence
  detection and adaptive demodulation technique using the {CNN} for the
  {OAM-FSO} communication,'' \emph{Optics Express}, vol.~26, no.~8, pp.
  10\,494--10\,508, Apr 2018.

\bibitem{EsmailOPEX21}
M.~A. Esmail, W.~S. Saif, A.~M. Ragheb, and S.~A. Alshebeili, ``Free space
  optic channel monitoring using machine learning,'' \emph{Optics Express},
  vol.~29, no.~7, pp. 10\,967--10\,981, Mar 2021.

\bibitem{GANGoodfellow}
I.~Goodfellow, J.~Pouget-Abadie, M.~Mirza, B.~Xu, D.~Warde-Farley, S.~Ozair,
  A.~Courville, and Y.~Bengio, ``Generative adversarial networks,'' in
  \emph{Proceedings of the International Conference on Neural Information
  Processing Systems (NIPS)}, 2014, p. 2672–2680.

\bibitem{GANYang}
Y.~{Yang}, Y.~{Li}, W.~{Zhang}, F.~{Qin}, P.~{Zhu}, and C.~{Wang},
  ``Generative-adversarial-network-based wireless channel modeling: Challenges
  and opportunities,'' \emph{IEEE Communications Magazine}, vol.~57, no.~3, pp.
  22--27, 2019.

\bibitem{Stat1}
``Africa internet users, 2020 population and facebook statistics,''
  \url{https://www.internetworldstats.com/stats.htm}, last accessed:
  18-04-2021.

\bibitem{YaacoubProcIEEE}
E.~{Yaacoub} and M.~{Alouini}, ``A key {6G} challenge and
  opportunity—connecting the base of the pyramid: A survey on rural
  connectivity,'' \emph{Proceedings of the IEEE}, vol. 108, no.~4, pp.
  533--582, 2020.

\bibitem{YaacoubIoTMag}
E.~Yaacoub and M.-S. Alouini, ``Efficient fronthaul and backhaul connectivity
  for iot traffic in rural areas,'' \emph{IEEE Internet of Things Magazine},
  vol.~4, no.~1, pp. 60--66, 2021.

\bibitem{MMIMO}
E.~{Björnson}, J.~{Hoydis}, and L.~{Sanguinetti}, ``Massive mimo has unlimited
  capacity,'' \emph{IEEE Transactions on Wireless Communications}, vol.~17,
  no.~1, pp. 574--590, 2018.

\bibitem{AlkhazragiOL21}
O.~Alkhazragi, A.~Trichili, I.~Ashry, T.~K. Ng, M.-S. Alouini, and B.~S. Ooi,
  ``Wide-field-of-view optical detectors using fused fiber-optic tapers,''
  \emph{Optics Letters}, vol.~46, no.~8, pp. 1916--1919, Apr 2021.

\bibitem{WillnerAOP15}
A.~E. Willner, H.~Huang, Y.~Yan, Y.~Ren, N.~Ahmed, G.~Xie, C.~Bao, L.~Li,
  Y.~Cao, Z.~Zhao, J.~Wang, M.~P.~J. Lavery, M.~Tur, S.~Ramachandran, A.~F.
  Molisch, N.~Ashrafi, and S.~Ashrafi, ``Optical communications using orbital
  angular momentum beams,'' \emph{Advances in Optics and Photonics}, vol.~7,
  no.~1, pp. 66--106, 2015.

\bibitem{AllenOAM}
L.~Allen, M.~W. Beijersbergen, R.~J.~C. Spreeuw, and J.~P. Woerdman, ``Orbital
  angular momentum of light and the transformation of {L}aguerre-{G}aussian
  laser modes,'' \emph{Physical Review A}, vol.~45, pp. 8185--8189, 1992.

\bibitem{GibsonOE04}
G.~Gibson, J.~Courtial, M.~J. Padgett, M.~Vasnetsov, V.~Pas'ko, S.~M. Barnett,
  and S.~Franke-Arnold, ``Free-space information transfer using light beams
  carrying orbital angular momentum,'' \emph{Optics Express}, vol.~12, no.~22,
  pp. 5448--5456, Nov 2004.

\bibitem{Tamburini2012}
F.~Tamburini, E.~Mari, A.~Sponselli, B.~Thid{\'{e}}, A.~Bianchini, and
  F.~Romanato, ``Encoding many channels on the same frequency through radio
  vorticity: first experimental test,'' \emph{New Journal of Physics}, vol.~14,
  no.~3, p. 033001, 2012.

\bibitem{JWangTbpsFSO}
J.~Wang, J.-Y. Yang, I.~M. Fazal, N.~Ahmed, Y.~Yan, H.~Huang, Y.~Ren, Y.~Yue,
  S.~Dolinar, M.~Tur, and A.~E. Willner, ``Terabit free-space data transmission
  employing orbital angular momentum multiplexing,'' \emph{Nature Photonics},
  vol.~6, no.~7, pp. 488--496, 2012.

\bibitem{100TbpsOAM}
H.~Huang, G.~Xie, Y.~Yan, N.~Ahmed, Y.~Ren, Y.~Yue, D.~Rogawski, M.~J. Willner,
  B.~I. Erkmen, K.~M. Birnbaum, S.~J. Dolinar, M.~P.~J. Lavery, M.~J. Padgett,
  M.~Tur, and A.~E. Willner, ``100 {Tbit/s} free-space data link enabled by
  three-dimensional multiplexing of orbital angular momentum, polarization, and
  wavelength,'' \emph{Optics Letters}, vol.~39, no.~2, pp. 197--200, 2014.

\bibitem{MMWOAM}
Y.~Yan, G.~Xie, M.~P. Lavery, H.~Huang, N.~Ahmed, C.~Bao, Y.~Ren, Y.~Cao,
  L.~Li, Z.~Zhao, A.~F. Molisch, M.~Tur, M.~J. Padgett, and A.~E. Willner,
  ``High-capacity millimetre-wave communications with orbital angular momentum
  multiplexing,'' \emph{Nature Communications}, vol.~5, no.~1, pp. 1--9, 2014.

\bibitem{cox2018diversity}
M.~A. Cox, L.~Cheng, C.~Rosales-Guzm{\'{a}}n, and A.~Forbes, ``{Modal Diversity
  for Robust Free-Space Optical Communications},'' \emph{Physical Review
  Applied}, vol.~10, no.~2, p. 024020, 2018.

\bibitem{OAMFSOTHz}
I.~B. {Djordjevic}, ``{OAM}-based hybrid free-space optical-terahertz
  multidimensional coded modulation and physical-layer security,'' \emph{IEEE
  Photonics Journal}, vol.~9, no.~4, pp. 1--12, 2017.

\bibitem{ElayanTHz}
H.~{Elayan}, O.~{Amin}, R.~M. {Shubair}, and M.-S. {Alouini}, ``Terahertz
  communication: The opportunities of wireless technology beyond {5G},'' in
  \emph{2018 International Conference on Advanced Communication Technologies
  and Networking (CommNet)}, 2018, pp. 1--5.

\bibitem{THzNatElec18}
K.~Sengupta, T.~Nagatsuma, and D.~M. Mittleman, ``Terahertz integrated
  electronic and hybrid electronic–photonic systems,'' \emph{Nature
  Electronics}, vol.~1, p. 622–635, 2018.

\bibitem{HadiSurveyTHz}
H.~Sarieddeen, M.-S. Alouini, and T.~Y. Al-Naffouri, ``An overview of signal
  processing techniques for terahertz communications,'' \emph{arXiv preprint},
  2020, arXiv:2005.13176.

\bibitem{THzResilience}
K.~Su, L.~Moeller, R.~B. Barat, and J.~F. Federici, ``Experimental comparison
  of terahertz and infrared data signal attenuation in dust clouds,''
  \emph{Journal of the Optical Society of America A}, vol.~29, no.~11, pp.
  2360--2366, Nov 2012.

\bibitem{THzSatellite}
J.~Y. {Suen}, M.~T. {Fang}, S.~P. {Denny}, and P.~M. {Lubin}, ``Modeling of
  terabit geostationary terahertz satellite links from globally dry
  locations,'' \emph{IEEE Transactions on Terahertz Science and Technology},
  vol.~5, no.~2, pp. 299--313, 2015.

\bibitem{SolarPoweredRF}
V.~{Chamola} and B.~{Sikdar}, ``Solar powered cellular base stations: current
  scenario, issues and proposed solutions,'' \emph{IEEE Communications
  Magazine}, vol.~54, no.~5, pp. 108--114, 2016.

\bibitem{HaasEnergies}
S.~Das, E.~Poves, J.~Fakidis, A.~Sparks, S.~Videv, and H.~Haas., ``Towards
  energy neutral wireless communications: Photovoltaic cells to connect remote
  areas,'' \emph{Energies}, vol.~12, no. 3772, 2019.

\bibitem{HaasGaAs}
J.~{Fakidis}, H.~{Helmers}, and H.~{Haas}, ``Simultaneous wireless data and
  power transfer for a {1-Gb/s GaAs VCSEL} and photovoltaic link,'' \emph{IEEE
  Photonics Technology Letters}, vol.~32, no.~19, pp. 1277--1280, 2020.

\bibitem{MIMOSLIPT}
I.~Tavakkolnia, L.~K. Jagadamma, R.~Bian, P.~P. Manousiadis, S.~Videv, G.~A.
  Turnbull, I.~D.~W. Samuel, and H.~Haas, ``Organic photovoltaics for
  simultaneous energy harvesting and high-speed {MIMO} optical wireless
  communications,'' \emph{Light: Science \& Application}, vol.~10, no.~41,
  2021.

\end{thebibliography}

\end{document}